\documentclass[useAMS, usenatbib, usegraphicx]{mn2e}
\usepackage[utf8]{inputenc}
\usepackage[T1]{fontenc}
\usepackage{color,times,journals}
\definecolor{green}{rgb}{0,0.5,0}
\definecolor{grey}{rgb}{0.4,0.5,0.7}

\usepackage{color}

\def\be{\begin{equation}}
\def\ee{\end{equation}}
\def\bea{\begin{eqnarray}}
\def\eea{\end{eqnarray}}

\newcommand{\PR}[1]{\ensuremath{\left[#1\right]}}
\newcommand{\PC}[1]{\ensuremath{\left(#1\right)}}

\title[Anisotropic $q$-Gaussian velocity distributions] 
{Anisotropic $q$-Gaussian 3D velocity distributions in $\Lambda$CDM haloes}
\author[]
{Leandro Beraldo e Silva$^{1,2,3}$\thanks{E-mail: lberaldo@if.usp.br}, 
Gary A. Mamon$^1$\thanks{E-mail: gam@iap.fr},
Manuel Duarte$^1$,
\newauthor
Rados{\l}aw Wojtak$^{4,5}$,
S\'ebastien Peirani$^1$
\& Gwena\"el Bou\'e$^{6,1}$
\\ 
$^{1}$ Institut d'Astrophysique de Paris (UMR 7095: CNRS \&
UPMC -- Sorbonne Universit\'es), F-75014 Paris, France\\
$^{2}$ Departamento de F\'isica Matem\'atica, Instituto de F\'{\i}sica,
Universidade de S\~ao Paulo, S\~ao Paulo SP, Brazil\\
$^{3}$ CAPES Foundation, Ministry of Education of Brazil, Bras\'ilia - DF
70.040-020, Brazil\\
$^4$ Kavli Institute for Particle Astrophysics and Cosmology, Stanford
University, Menlo Park CA, USA\\
$^5$ Dark Cosmology Centre, Niels Bohr Institute, University of Copenhagen, Denmark\\
$^6$ Institut de M\'ecanique C\'eleste et de Calcul des \'Eph\'emerides  (UMR
8028: CNRS \& UPMC -- Sorbonne Universit\'es), Observatoire de Paris, F-75014
Paris, France}

\begin{document}
%
\date{\today}
\pagerange{\pageref{firstpage}--\pageref{lastpage}} \pubyear{2013}
\maketitle
\label{firstpage}
%
\begin{abstract}
The velocity distribution function (VDF) of dark matter (DM) haloes in
$\Lambda$CDM dissipationless cosmological simulations, which must be
non-separable in its radial and tangential components, is still poorly
known. We present the first single-parameter, non-separable, anisotropic
model for the VDF in $\Lambda$CDM haloes, built from an isotropic $q$-Gaussian
(Tsallis) VDF of the isotropic set of dimensionless spherical velocity
components (after subtraction of streaming motions), normalized by the
respective velocity dispersions. We test our VDF on 90 cluster-mass haloes of a
dissipationless cosmological simulation.

Beyond the virial radius, $r_{\rm vir}$, 
our model VDF adequately reproduces that measured in the simulated
haloes, but no $q$-Gaussian model can adequately represent the VDF within
$r_{\rm vir}$, as the speed distribution function is then
flatter-topped than any $q$-Gaussian can allow. Nevertheless, our VDF fits
significantly better the simulations than the commonly used Maxwellian
(Gaussian) distribution, at virtually all radii within $5\,r_{\rm vir}$. Within 0.4
(1) $r_{\rm vir}$, the non-Gaussianity index $q$ is (roughly) linearly
related to the slope of the density profile and also to the velocity
anisotropy profile. We provide a parametrization of the modulation of $q$
with radius for both the median fits and the fit of the stacked halo. At
radii of a few percent of $r_{\rm vir}$, corresponding to the Solar position
in the Milky Way, our best-fit VDF, although fitting better the simulations
than the Gaussian one, overproduces significantly the fraction of high
velocity objects, indicating that one should not blindly use these
$q$-Gaussian fits to make predictions on the direct detection rate of DM
particles.

\end{abstract}
%
%
%

\section{Introduction}
\label{sec_intro}
While dark matter appears to constitute 85\% of the mass of the Universe,
much work is being performed to detect dark matter particles and to quantify
its distribution in astronomical systems.
In particular, experiments have been developed in order to detect the
passage of dark matter particles through terrestrial detectors: DAMA
\citep{Bernabei+13}, CoGeNT \citep{Aalseth+13},
CRESST-II \citep{Petricca+12}, CDMS-Si \citep{Agnese+13}, and Xenon100 \citep{Aprile+12}.
The knowledge of the precise high-end part of the distribution of space (3D)
velocities (hereafter velocity distribution function or VDF) in the inner
halo, corresponding to the Solar position in the Milky Way galaxy, 
is required to quantify the expected event rate in
direct dark matter detection experiments.
Indeed, these experiments involve a 
detection threshold in kinetic energy, which for light (e.g. $\approx 10
\,\rm GeV$ in mass) dark matter particles
corresponds to velocities of order of $300 \, \rm km \,
s^{-1}$, i.e. somewhat higher than the expected
velocity dispersion of halo dark matter particles in the Solar neighbourhood.
With this goal in mind, the  VDF
in $\Lambda$ cold dark matter ($\Lambda$CDM) haloes has drawn attention during the last few
  years \citep{FS09,Vogelsberger+09,LNAT10,Kuhlen_2010,LSWW11,Mao+13,Pato_2013}. 

The knowledge of the VDF is also
important 
for modeling the radial profiles of mass (including dark matter) and velocity
anisotropy of quasi-spherical systems from the distribution of their tracers
(stars in galaxies; galaxies in clusters) in projected phase space (PPS:
projected radius and line-of-sight velocity).
The cleanest way to perform this mass / anisotropy analysis is to model the
distribution of tracers in PPS, but this requires a triple integral 
of the six-dimensional distribution function (DF) expressed
in terms of energy and angular momentum, $f(E,J)$ \citep{DM92}.
For example, the method of  \cite{Wojtak+09} that
starts from the $\Lambda$CDM halo DF of \cite{Wojtak+08}
is very slow (requiring a day on a single processor 
to run for a 500-tracer system with full error sampling  from
Markov Chain Monte-Carlo (MCMC) methods). Orbit modeling
\citep{Schwarzschild79,RT84,ST96} is much slower, thus preventing proper
error sampling by MCMC.
Recently, \cite*{Mamon+13} have developed an algorithm called MAMPOSSt, in which the distribution of tracers in
PPS is expressed as a single  integral: 
$f(E,J)$ is replaced by the distribution of line-of-sight velocities at a
given (3D) position, which in turn depends on
 the combination of the radial profiles
of the total mass and the  velocity
anisotropy 
\begin{equation}
\beta = 1-{\sigma_\theta^2+\sigma_\phi^2 \over 2\,\sigma_r^2} \ ,
\label{betadef}
\end{equation}
\emph{as well as} 
a suitably simple form for the VDF.
%
So far, MAMPOSSt has only been used with a Maxwellian (or Gaussian\footnote{We will refer to the
  \emph{Maxwellian} VDF in the physical context, and to the
  \emph{Gaussian} VDF in the mathematical context.}) VDF
(\citealp{Mamon+13,Biviano+13}; \citealp*{Munari+14}; \citealp{Guennou+14}; \citealp{Mamon_Oxford+15}).
However, $N$-body simulations (both cosmological and academic ones) indicate
that VDFs show 
departures from Gaussianity in their radial component \citep{WLGM05,HMZS06}
and also their tangential component \citep{HMZS06}.
Moreover, forcing Gaussianity in the VDF of isotropic systems built from the Jeans equation of
local dynamical equilibrium leads to unstable density profiles, whereas
analogous systems built from distribution
functions are stable \citep{KMM04}.

The appropriate statistical mechanical description of the 6D
structure of  self-gravitating
spherical systems is an old and still open problem.  
A generalization of the
Maxwellian VDF has been proposed in the context of the
non-extensive thermodynamics developed by \cite{Tsallis88}.
The Tsallis VDF, alternatively called $q$-Gaussian, 
is
equivalent to the polytropic gas model \citep[see][]{Plastino_Plastino_1993},
of which the isothermal sphere is a particular case. 
The Tsallis VDF has been applied to
describe phenomena of diverse fields of physics, particularly
self-gravitating systems, but also to the direct detection of dark matter
particles \citep*{VHH08}.

To the best of our
knowledge, all analyses of non-Maxwellian VDFs, with one exception,
assume
velocity isotropy \citep{Vogelsberger+09,LSWW11,Mao+13},
a VDF that is
separable into its radial and tangential components
\citep{HMZS06,FS09,Kuhlen_2010}, or tried both \citep{LNAT10}.
Recently, \cite{Hunter14} generalized the VDF of \cite{Mao+13} to a joint form of
radial and tangential velocities, but his model involves 3 parameters.

Interestingly, the radial and tangential  components of the  VDF of
structures in cosmological and academic $N$-body simulations are well fit, separately, by
the $q$-Gaussian formula \citep{HMZS06}, although other modifications to the
Gaussian have been shown to also fit well the VDFs of haloes (at the solar
radius, \citealp{FS09,Kuhlen_2010,LSWW11,Mao+13}).
Moreover, simulations of both collapsing structures and cosmological haloes indicate 
that  for both for the radial velocity and the tangential velocity
distributions,
the $q$ parameter of non-Gaussianity is found to vary roughly linearly with
the slope of the density profile for radii where the slopes are
$\gamma={\rm d}\ln\rho/{\rm d}\ln r$ between --2.7 and --1
\citep{HMZS06}. 
Finally, \cite{HS12} demonstrate that the tangential VDF must scale, outside
its wings, as
$[1+v^2/(3\sigma_v^2)]^{-5/2}$ at all radii.

In fact,
\emph{were} the dynamical evolution of
these systems just determined by two-body interactions, as
is the case for ideal gases, i.e., if the two-body relaxation time were
short, then the system 
would rapidly evolve to
isotropic velocities in a short time scale, the
distribution function would then depend
solely on energy, $f=f(E)$, and could be obtained from the density profile
\citep{Eddington1916}, and finally, the velocity modulus distribution function at
radius $r$ would then simply be $f_v(v|r) \propto v^2 f(v^2/2+\Phi(r))$.
%
However, 
in most large-scale astronomical systems (galaxies and clusters),
the two-body relaxation time of the dark matter component is longer than
the age of the Universe. One might still expect that violent
relaxation, caused by a rapidly varying gravitational potential
\citep{LyndenBell67}, will redistribute energies and lead to a possibly
stationary configuration. However, violent relaxation is not thought to be
long-term, 
and, furthermore, the energies are not completely redistributed
(\citealp*{Madsen87,KMS93}; \citealp{BeraldoeSilva+14}).

On the other hand, simulations and observational modeling suggest that the VDF in elliptical galaxies and
galaxy clusters is most likely anisotropic. Indeed, $\Lambda$CDM haloes of
cluster-mass haloes show
radial velocities at outer (e.g., \citealp{Lemze+12}) or all \citep*{WGK13}
radii.
Moreover, dynamical studies of galaxies \citep{Wojtak&Mamon13} and clusters
\citep{BK04,Lokas+06,WL10,Biviano+13,Munari+14}
point to radial outer
velocity anisotropy.

In principle, the VDF can be deduced from the DF.
For anisotropic spherical systems,
since $\int\!\!\int 2\pi\,v_{\rm t}\,f(E,J)\,{\rm d}v_r\,{\rm d}v_{\rm t} =
\rho$, the VDF at radius $r$ will be
\begin{equation}
f_{\rm v}(v_r,v_{\rm t}|r) = {2\pi\over \rho(r)}\,v_{\rm t}\,f\left([v_r^2+v_{\rm
    t}^2]/2+\Phi(r),r\,v_{\rm t}\right) \ .
\label{eq_vdf_from_df}
\end{equation}
For example, one could use the separable form of the DF that \cite{Wojtak+08}
measured for $\Lambda$CDM haloes. However, that DF involves a total of 8
parameters, so, although interesting, the approach of
equation~(\ref{eq_vdf_from_df})  
is left for future work (see also \citealp{FG14}).
%

The separability of the DF in energy and angular momentum thus indicates that
the VDF of $\Lambda$CDM haloes
is a non-separable function of radial and
tangential velocities.
Indeed, if $f(E,J) = f_E(E)\,f_J(J)$, then according to
equation~(\ref{eq_vdf_from_df}), $v_{\rm t}$ cannot be separated from $v_r$
within $f_E(E)$, unless $f_E(E) = {\rm cst}\,\exp(-a E)$, where $a$ is a
constant, but this is not the case
for $\Lambda$CDM haloes \citep{Wojtak+08}.





Thanks to the interest in direct dark matter detection, 
most work on the VDF has been restricted to radii of $\approx
3$ percent of the virial radius, $r_{\rm vir}$,
i.e. the position of the Earth in the halo of the Milky
Way.
On the other hand, as in many other mass / velocity anisotropy modeling methods, MAMPOSSt involves
integrals along the line-of-sight (LOS), corresponding to physical radii
extending from $r=R$ to infinity, in principle. In practice, the Hubble flow
stretches the velocity vs. distance-to-halo-centre relation so that beyond
$r_{\rm max} \approx 13\, r_{\rm vir}$, the line-of-sight velocities extend beyond
$3\,\sigma_{\rm LOS}$ (e.g., \citealp*{Mamon+10}). 
Thus,
 the knowledge of the VDF is required at all radii from the halo
centre to $\approx 13 \,r_{\rm vir}$.

In this work, we propose the first anisotropic VDF for $\Lambda$CDM haloes 
that is a non-separable function of
radial and tangential velocities, after that of \cite{Hunter14}. Our VDF is
 an extension of the $q$-Gaussian VDF to spherical
systems with anisotropic velocities. Instead of 3 parameters as in the
\citeauthor{Hunter14} VDF, ours has only one parameter: the non-Gaussianity
index $q$. We do not advocate any fundamental basis
for the $q$-Gaussian velocity distribution. Instead, we treat it as a
powerful parametrization that allows us to phenomenologically describe
systems whose 3D velocity distributions depart from the Gaussian
distribution. We then fit our anisotropic $q$-Gaussian model to the VDF of
simulated $\Lambda$CDM haloes, between 0.03 and $13 \,r_{\rm vir}$, to check if it provides a significantly better
representation of the VDF than does the Gaussian model with
one parameter less.

Note that this non-separable form of the VDF has the practical advantage that
it is straightforward to compute the 
distribution of line-of-sight velocities from it, while the distribution of
line-of-sight velocities for a separable $q$-Gaussian VDF cannot be expressed
in analytical form in a single quadrature. This means that this non-separable
$q$-Gaussian VDF can be incorporated into the MAMPOSSt mass/orbit modeling technique.

In Sect.~\ref{sec_maxwell}, we review the classical and simplest case, of the
Gaussian velocity distribution from the Maxwellian approach. In
Sect.~\ref{sec_tsallis}, 
we 
present the $q$-Gaussian velocity distribution and briefly describe its
extensions to thermodynamics.
Then, in Sect.~\ref{sec_anisotropic_tsallis}, we generalize the $q$-Gaussian
velocity distribution to spherical systems with anisotropic velocities. 
In Sect.~\ref{sec_simulations}, we describe the simulated data that we use,
while in Sect.~\ref{sec_fitting} we explain how we arrange the data and fit
the non-Gaussianity index. 
In Sect.~\ref{sec_results}, we analyze the properties of the 3D velocity distribution of
cluster-mass $\Lambda$CDM haloes as a function of radial distance to the halo
centre.
We discuss our results in Sect.~\ref{sec_discuss}.


\section{Gaussian velocity distribution}
\label{sec_maxwell}
The velocity distribution of an ideal gas in equilibrium was first determined
by \cite{Maxwell1860}, based on two symmetry hypotheses
(see \citealp*{Sommerfeld_1993,Silva_Plastino_Lima_1998}; \citealp{DGLR07}):
\begin{enumerate}
\item The velocity distribution $F\PC{\bmath{v}}$ is isotropic. This implies that $F\PC{\bmath{v}} =
  F\PC{\sqrt{\bmath{v}\cdot\bmath{v}}}=F\PC{v}$.
\item The 3 directions are statistically independent. This, using cartesian
  coordinates, implies that $F\PC{\bmath{v}} =
  f_1\PC{v_x}f_2\PC{v_y}f_3\PC{v_z}$, where $f_1$, $f_2$ and $f_3$ can be
  different in general.  
\end{enumerate}
Together, these hypotheses imply that
\be
F\PC{v} =f\PC{v_x}f\PC{v_y}f\PC{v_z}.
\label{eq_joint_Maxwell_general}
\ee
Following standard steps, from eq.~(\ref{eq_joint_Maxwell_general}) we write
\be
\ln F\PC{v} =\ln f\PC{v_x} + \ln f\PC{v_y} + \ln f\PC{v_z}
\label{eq_lnF_Maxwell}
\ee
and differentiate both sides relative to $v_x$, obtaining
\be 
\frac{1}{v}\,\frac{{\rm d}\ln F}{{\rm d}v} = \frac{1}{v_x}\,\frac{{\rm d}\ln f}{{\rm d}v_x} \ .
\label{eq_diffvx}
\ee
In equation~(\ref{eq_diffvx}), the left-hand side is only a function of $v$,
while the right-hand side is only a function of $v_x$, which implies that
both are equal to some constant
$-k$. This leads to
\be
f\PC{v_x} \propto \exp \left(-\frac{k}{2}v_x^2 \right) \ .
\label{eq_vxsol}
\ee
Equation~(\ref{eq_vxsol}) also holds for $v_y$ and $v_z$. Therefore, 
the joint velocity distribution is
\be
F(\bmath{v})  = F\PC{v} = A \exp \left({-\frac{k}{2}v^2}\right)\ ,
\label{eq_vdisMaxwell}
\ee
where $A = [k/(2\pi)]^{3/2}$ is determined by the normalization condition
\be
\int F\PC{\bmath{v}}{\rm d}^3\bmath{v} = \int_0^\infty F(v)\, 4 \pi v^2\,{\rm d}v = 1 \ ,
\label{eq_normal_cond}
\ee
while $k = 1/\sigma_v^2$, with $\sigma_v$ the one-dimensional velocity
dispersion determined by the 2nd velocity moment condition
\begin{equation} 
\int F(\bmath{v})\,v^2\,{\rm d}^3\bmath{v} = \int_0^\infty F(v)\, 4 \pi v^4\,{\rm d}v = 3\,\sigma_v^2
\ .
\label{eq_normal_cond2}
\end{equation}

\cite{Boltzmann1872}
showed that the velocity distribution of equation~(\ref{eq_vdisMaxwell})  is not
changed by molecular collisions and obtained the expression for the entropy
that, maximized, gives the velocity distribution previously derived by
Maxwell.

\section{Tsallis (or \lowercase{q}-gaussian) velocity distribution}
\label{sec_tsallis}

The DF of isotropic spherical systems 
implied by equation~(\ref{eq_vdisMaxwell}) is
then $f(E) \propto \exp(-E/\sigma^2)$.
Since
the joint assumptions of isotropy (i) and separability of the velocity
components (ii) lead
to
purely exponential energy distributions, then non-exponential energy
distributions of isotropic systems will necessarily lead to the
non-separability of the VDF. 
For example, if the DF is truncated because of escaping particles (e.g.,
\citealp{King66}),
the VDF will be non-separable.
This provides a natural motivation to explore
non-Maxwellian VDFs such as the Tsallis distribution.

Historically speaking, the $q$-Gaussian velocity distribution was derived in
the opposite order. Firstly, a generalized version of the Boltzmann entropy
was proposed by \cite{Tsallis88}, and then the VDF was obtained by
maximizing this Tsallis entropy 
\citep{Plastino_Plastino_1993}. 
Finally, the same velocity distribution was obtained \citep{Silva_Plastino_Lima_1998} following symmetry arguments similar to that of Maxwell.

In fact, assuming the velocity
isotropy hypothesis (i) above --- which allows us to write $F\PC{\bmath{v}}  =
F\PC{v}$ --- but abandoning the coordinate-independence hypothesis (ii),
\cite{Silva_Plastino_Lima_1998} proposed, as a generalization of the joint
Maxwellian distribution (equation~\ref{eq_joint_Maxwell_general}), 
the expression
\be
F\PC{v}=\exp_q\PR{\displaystyle\sum\limits_{i=x}^z f^{q-1}\PC{v_i}\ln_q f\PC{v_i}},
\label{eq_FofvTsallis}
\ee
where
\[
\exp_q\PC{f} = \PR{1 + (1-q)f}^{1/(1-q)}
\]
is called the $q$-exp function, and follows $\exp_q(f)\to {\rm e}^f$ as $q\to 1$,
while
\[
\ln_q\PC{f} = \frac{f^{1-q} - 1}{1-q} 
\]
is called the $q$-log function, and follows $\ln_q(f)\to \ln f$ as $q\to1$.
One can easily check that $\exp_q[\ln_q(f)]=f$.
Then, in the limit $q\to1$,
the joint velocity distribution of equation~(\ref{eq_FofvTsallis})  
tends to the Maxwellian
distribution. Following the same steps as for the Maxwellian VDF, we arrive at
\be
F\PC{\bmath{v}}  = F\PC{v} = B_q \PR{1 - (1-q)\frac{k}{2}v^2}^{1/(1-q)},
\label{eq_joint_Tsallis}
\ee
where the constants $B_q$ and $k$ are obtained using the normalization
equations~(\ref{eq_normal_cond}) and (\ref{eq_normal_cond2})
(see \citealp{Silva_Alcaniz_2003})
\begin{equation} 
B_q = 
\left ({k\over 2\pi}\right)^{3/2}
\!\!\left\{
\!\!\!\!
\begin{array}{ll}
\displaystyle
(1\!-\!q)^{3/2} \,
{\Gamma[1/(1\!-\!q)+5/2]\over \Gamma[1/(1\!-\!q)+1]}
&\!\!(0\!<\!q\!<\!1) \ , \\
&  \\
\displaystyle
(q\!-\!1)^{3/2} {\Gamma[1/(q\!-\!1)]\over \Gamma[1/(q\!-\!1)-3/2] } 
&\!\!(1\!<\!q\!<\!5/3) \ ,
\end{array} 
\right. 
\!\!\!\!\!\!\!\!\!\!\!\!\!\!\!\!\!\!
\label{eq_joint_Tsallis_u}
\end{equation}
where\footnote{This value of $k$ is different from what would be inferred from eq.~(7) of
    \cite{Silva_Alcaniz_2003}, 
 who compute the second velocity moment as $\int v^2 F^q(\bmath{v}) \,{\rm d}^3\bmath{v}$
 instead of as in the left-hand-side of equation~(\ref{eq_normal_cond2}). 
We
 are using the $q$-Gaussian VDF of equation~(\ref{eq_joint_Tsallis}) 
as an empirical model and are not considering the effects of the
non-extensive thermodynamics proposed by 
\cite{Tsallis88}, hence our use of a classical 2nd velocity moment to
derive $k$.
}
\begin{equation}
k = {2\over 7-5q}\, {1\over \sigma_v^2}\ .
\label{kofq}
\end{equation}
This value of $k$ imposes the tighter restriction $q<7/5$.
Note the velocity limit of $\sqrt{2/[k(1-q)]}$ when $q<1$.
The shape of $F(v)$ depends on the value of $q$.
While $F(v)$ is Maxwellian for $q=1$, it has a flatter top and is sharply
truncated for $q<1$ and has a cuspier top with wider wings when $q>1$.

\section{Anisotropic Tsallis (\lowercase{\emph{q}}-gaussian) velocity distribution}
\label{sec_anisotropic_tsallis}

The velocity distribution of equation~(\ref{eq_joint_Tsallis})
depends only on the modulus of the velocity, hence is isotropic,
as expected by construction.
Since simulated astrophysical systems have anisotropic velocities, we now
extend the $q$-Gaussian velocity distribution to anisotropic velocities.


One possible approach to extend the $q$-Gaussian VDF to
anisotropic velocities would be to maximize the Tsallis entropy, as done by
\cite{Plastino_Plastino_1993}, but with additional constraints
\citep[see][]{Stiavelli_Bertin_1987}.
Instead,
inspired by eq.~(\ref{eq_joint_Tsallis}), and concerned with spherically
symmetric self-gravitating systems, 
and correcting for streaming motions (e.g., streaming radial motions beyond the
virial radius), 
we introduce the anisotropic $q$-Gaussian
VDF as
\begin{eqnarray}
F\PC{\bmath{v}}&\!\!\!\!=\!\!\!\!& {C_q\over (1-\beta)\,\sigma_r^3}\nonumber \\
&\!\!\!\!\mbox{}\!\!\!\!& \times  
\left[1 - (1\!-\!q){D_q\over2}\right. \nonumber \\
&\!\!\!\!\mbox{}\!\!\!\!& \times  
\quad \left.
\left(\frac{\left(v_r\!-\!\overline{v_r}\right)^2}{\sigma_r^2} 
  \!+\!   \frac{\left(v_\theta\!-\!\overline{v_\theta}\right)^2}{\sigma_\theta^2} 
  \!+\! \frac{\left(v_\phi\!-\!\overline{v_\phi}\right)^2}{\sigma_\phi^2}\right)
\right]^{1/(1\!-\!q)} \!\!\!\!\!\!\!\!\!\!\!\!\!\!\!\!,
\label{eq_anisotropic_F_general}
\end{eqnarray}
where the $\overline{v_i}$ and $\sigma_i$ are respectively the mean streaming velocities
and velocity dispersions 
in the direction $i$
of the spherical coordinate system, while $C_q$ and $D_q$ are constants
(dependent on $q$) that we shall determine below.

Defining the \emph{dimensionless normalized velocities} as 
\begin{equation}
u_i = {v_i - \overline{v_i}\over \sigma_i} \ ,
\label{uofi}
\end{equation}
and noting that
the Jacobian relating the $v_i$ to the $u_i$ is $(1-\beta)\,\sigma_r^3$,
equation~(\ref{eq_anisotropic_F_general}) can equivalently be written
\begin{equation}
F(\bmath{u}) = C_q \PR{1 -
  (1\!-\!q){D_q\over2}\PC{u_r^2+u_\theta^2+u_\phi^2}}^{1/(1\!-\!q)} \ .
\label{eq_anisotropic_F_u}
\end{equation}
One notices that the vector field $\bmath u$ is isotropic by construction.
Equations~(\ref{eq_normal_cond}) and (\ref{eq_normal_cond2}) become
\begin{eqnarray}
\int F(\bmath{u})\,{\rm d}^3\bmath{u} &=& \int_0^\infty F(u)\, 4 \pi
u^2\,{\rm d}u = 1 \ ,
\label{eq_normal_condu}
\\
\int F(\bmath{u})\,u^2\,{\rm d}^3\bmath{u} &=& \int_0^\infty F(u)\, 4 \pi
u^4\,{\rm d}u = 3 \ .
\label{eq_normal_cond2u}
\end{eqnarray}
Equation~(\ref{eq_anisotropic_F_u}) is identical to
equation~(\ref{eq_joint_Tsallis}) with $k$ taken from equation~(\ref{kofq}),
once one sets $\sigma_v$ in the latter 
equation to unity.
Therefore,
\begin{equation} 
D_q = {2\over 7 - 5\,q}
\label{eq_Dq}
\end{equation}
and
\begin{eqnarray}
C_q 
&\!\!\!\!=\!\!\!\!& \left ({D_q\over 2\pi}\right)^{3/2}\,
\nonumber \\
&\!\!\!\!\mbox{}\!\!\!\!& \quad \times
\left\{
\!\!\!\!
\begin{array}{ll}
\displaystyle
(1\!-\!q)^{3/2} \,
{\Gamma[1/(1\!-\!q)+5/2]\over \Gamma[1/(1\!-\!q)+1]}
&\!\!(0\!<\!q\!<\!1) \ , \\
&  \\
\displaystyle
(q\!-\!1)^{3/2} {\Gamma[1/(q\!-\!1)]\over \Gamma[1/(q\!-\!1)-3/2] } 
&\!\!(1\!<\!q\!<\!7/5) \ .
\end{array} 
\right. 
\!\!\!\!\!\!\!\!\!\!\!\!\!\!\!\!\!\!
\label{eq_Cq}
\end{eqnarray}
Considering the radial and tangential dimensionless normalized velocities
\begin{eqnarray}
u_r &=& {v_r-\overline{v_r}\over \sigma_r} \ , 
\label{eq_u_r} \\
u_{\rm t} &=& \sqrt{u_\theta^2+u_\phi^2} =
\sqrt{\left({v_\theta-\overline{v_\theta}\over \sigma_\theta}\right)^2
+\left({v_\phi-\overline{v_\phi}\over \sigma_\phi}\right)^2} \ ,
\label{eq_u_t}
\end{eqnarray} 
the probability distribution function of $(u_r,u_{\rm t})$ is then
\begin{eqnarray} 
F(u_r,u_{\rm t}|q) &\!\!=\!\!& 2\pi\, u_t \,F(u_r,u_\theta,u_\phi) \nonumber \\
&\!\!=\!\!&
2\pi\, C_q\,u_{\rm t}
\left[1-(1\!-\!q){D_q\over
    2}\!\left(u_r^2\!+\!u_{\rm
    t}^2\right)\right]^{1/(1\!-\!q)} \!\!\!\!\!\! \ .
\label{eq_anisotropic_F_ur_ut}
\end{eqnarray}
The VDF expressed in
dimensionless normalized velocities in equation~(\ref{eq_anisotropic_F_ur_ut}) is clearly
not separable into two terms respectively depending on $u_r$ and on  $u_{\rm t}$.

\subsection{Velocity modulus (or speed) distribution}
As for the Maxwellian VDF, it is interesting to define the probability 
distribution function of the modulus of the velocity, i.e. the speed distribution
function (SDF) of the $q$-Gaussian VDF.
Defining the 
dimensionless normalized speed as
\be
u = \sqrt{u_r^2 + u_{\rm t}^2} \ ,
\label{eq_u}
\ee
the SDF is
\begin{eqnarray}
G(u|q) 
&=&
4 \pi u^2 F\left(u_r,u_\theta,u_\phi\right) \nonumber \\
&=&
4\pi\,C_q\,
u^2 \PR{1 -
  (1\!-\!q){D_q\over2} u^2}^{1/(1\!-\!q)} \ ,
\label{eq_anisotropic_G}
\end{eqnarray}
where equation~(\ref{eq_anisotropic_F_u}) is used for the second equality.
Again, we have a maximum  velocity for $q<1$, which is now
$u_{\rm max} = \sqrt{2/[D_q(1-q)]} = \sqrt{(7-5q)/(1-q)}$.

\section{Simulations}
\label{sec_simulations}
To test the performance of the VDF of
equation~(\ref{eq_anisotropic_F_general}), we have analyzed a cosmological dark matter  $N$-body
simulation performed with Gadget-2 \citep{Springel05}.
The simulation was run with  $512^3$ particles in a periodic box
of comoving size $L=100 \, h^{-1} \, \rm Mpc$, 
using a WMAP7 cosmology: $\Omega_{\rm m}=0.272$,
$\Omega_\Lambda=0.728$, $h=0.704$, $\sigma_8=0.807$. The particle mass is
$5.62\times 10^8 h^{-1} {\rm M}_\odot$.
The Plummer-equivalent force softening is 5\% of the mean inter-particle
distance and kept constant in comoving units. This amounts to $0.05\,L/512 =
9.8 \, h^{-1} \, \rm kpc$. Initial conditions have been
generated using the {\sc MPgrafic} code \citep{Prunet+08}.

Haloes were extracted  with {\sc HaloMaker} 2.0 using a Friends-of-Friends
technique \citep{DEFW85}, with linking length $b=0.2$ (in units of the mean interparticle separation).

We selected 90
haloes from the $z=0$ output of the cosmological simulation,
divided in 3 subsamples of comparable mass: the first
subsample contains the 30 most massive haloes, while the other two subsamples
contain  haloes with geometric
mean differing by 0.5 and by 1.0 dex from the geometric mean of the first
subsample (i.e., the haloes of mass rank $53-82$ and $221-250$). 

\begin{figure}
\centering
\includegraphics[width=\hsize]{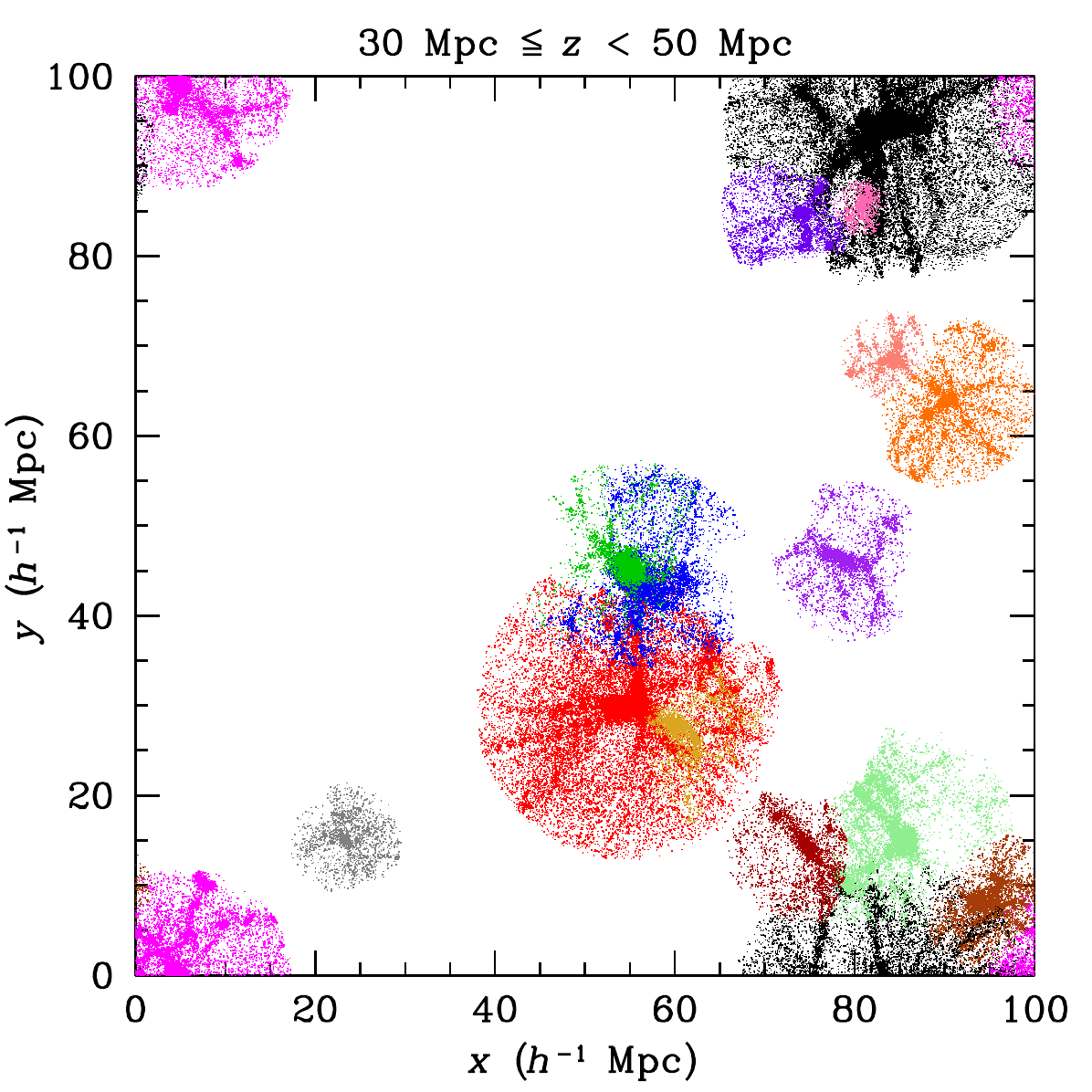} 
\caption{Illustration of the nearest assignment procedure on a $20 \, h^{-1} \, \rm
  Mpc$ slice of simulation box, with each of the 90 chosen haloes coded by a
  random colour (only 15 are present in this slice). 
The 160 other haloes among the top 250 are not shown, but one
  clearly sees their effects, i.e. on the purple halo 
(at $(80,46)\,h^{-1}\,\rm Mpc$).
Note that the periodic boundary conditions split the magenta halo
(at $(6,98)\,h^{-1}\,\rm Mpc$)
 into the
four corners, and the black halo
(at $(84,95)\,h^{-1}\,\rm Mpc$)
 between the top and bottom.
\label{slice}}
\end{figure}
We analyzed the haloes as follows (taking into account the periodic boundary
conditions at all steps). First, we refined the centre of each
of the 250 most massive haloes using an iterative median centre scheme, starting on the halo particles
returned by the halo finder, computing the median halo coordinates, and
restricting to the particles within half of the initial (virial) radius
around the new centre, iterating 3 times (each within smaller regions).
We then re-estimated the virial radii of the 250 most massive haloes by finding the
radius, $r_{100}$, where the mean density within the sphere centered on the newly determined centre
is 100 times the critical density of the
Universe at $z=0$.\footnote{For the cosmology of our simulation the mean
  density within the virial
  radius is 97 according to the approximation of \cite{BN98}.}
For this, we considered all radii out to 3 old virial radii from the newly
determined centre and solved $2 G M(r) / (H_0^2 r^3) = 100$ for $r$.

For our three subsamples of haloes, the median virial masses are then
$\langle M_{100}\rangle = 1.55\times 10^{14} M_\odot$, 
$6.10\times 10^{13} M_\odot$, 
and 
$1.99\times 10^{13} M_\odot$,
in each of the three subsamples, respectively.
The corresponding median virial radii are 
$\langle r_{100}\rangle = 1.39$, 1.02, and $0.70 \,\rm Mpc$, 
the softening length of the simulation is
0.010, 0.014, and 0.020 times these respective virial radii, and the median number
of particles within these virial radii are
$1.9\times10^5$, $7.6\times10^4$ and $2.5\times10^4$, respectively.

To avoid assigning particles outside of halo virial spheres to two or more haloes,
we reassigned all particles to the nearest of the 250 most massive haloes in units of their virial
radii.
Fig.~\ref{slice} illustrates the procedure. One notices that some haloes
are cut by other
haloes among the $250-90=160$ that are not shown, as
for example seen in the
left-lower-left part of the orange halo centered  near $(x,y)=(90,63) \, h^{-1} \, \rm Mpc$.

\section{Fitting procedure}
\label{sec_fitting}
We performed two types of  analyses. 
On one hand, we split each of the 90 haloes into radial bins of $\simeq 5000$
particles (where the radii of each halo were normalized to the virial radius of that halo), and then
normalized the 3 spherical velocity coordinates by subtracting the mean and dividing by the
dispersion, as in equation~(\ref{uofi}).
For each halo, we fit for $q$ vs. $r/r_{100}$ (see last paragraph of this section). 
We then performed linear interpolation (without extrapolation) of $q(r)$ and
other parameters on a grid of 27 geometrically-spaced radii,
from $\log r/r_{100}=-1.5$ to 1.1 in steps of 0.1,  i.e. from 
$0.03\,r_{100}$  (within which the definition of the centre and
the softening length of the simulation may affect the
results) to $13\,r_{100}$ (beyond which the Hubble flow moves the LOS velocities
beyond $\pm 3 \sigma_{\rm LOS}$, see Sect.~\ref{sec_intro}).
This allowed us  to
 determine the median values (over $\leq 90$ haloes) of $q$ in fixed radial bins.

On the other hand, we built a
\emph{stacked halo}, with all particles of our 90
haloes. 
We considered all 90 haloes together, using the normalized radii computed for
the individual haloes (see above). Within 175 radial bins of equal numbers
($\simeq 300\,000$) of particles, we normalized 
the 3 spherical velocity coordinates as for the individual haloes (again using
equation~\ref{uofi}). In the end, our stacked halo
contained 52$\,$351$\,$250 particles out to $13\,r_{100}$, with 
radii normalized by $r_{100}$ and dimensionless velocities $u_r$,
$u_\theta$ and $u_\phi$.
We then fit $q$ on either the 175 equal number radial bins or on the 27 geometrically spaced radial bins.

%

The maximum likelihood estimate of $q$ can be determined
from the distribution of $(u_r,u_{\rm t})$,
by minimizing
\begin{equation}
-\ln {\cal L}(q) = - \sum \ln F(u_r,u_{\rm t}|q) \ ,
\end{equation}
where $F(u_r,u_{\rm t}|q)$ is given in equation~(\ref{eq_anisotropic_F_ur_ut}) and
depends on $q$.
We have minimized, instead,
\begin{equation}
-\ln {\cal L'}(q) = - \sum \ln G(u|q) = - \ln {\cal L}(q) + {\rm
  extra\ term}\ ,
\label{eq_like2}
 \end{equation}
where $G(u|q)$ is given in equation~(\ref{eq_anisotropic_G}), and the extra
term is independent of $q$, hence minimizing $-\ln {\cal L'}$ is equivalent
to minimizing $-\ln {\cal L}$.
The minimization was performed using the simulated annealing method on the $\chi^2$ values.
 
\section{Results}
\label{sec_results}

\subsection{Radial profiles of non-Gaussianity}
\label{sec_profiles}
\begin{figure}
\includegraphics[width=\hsize,bb=0 140 600 700]{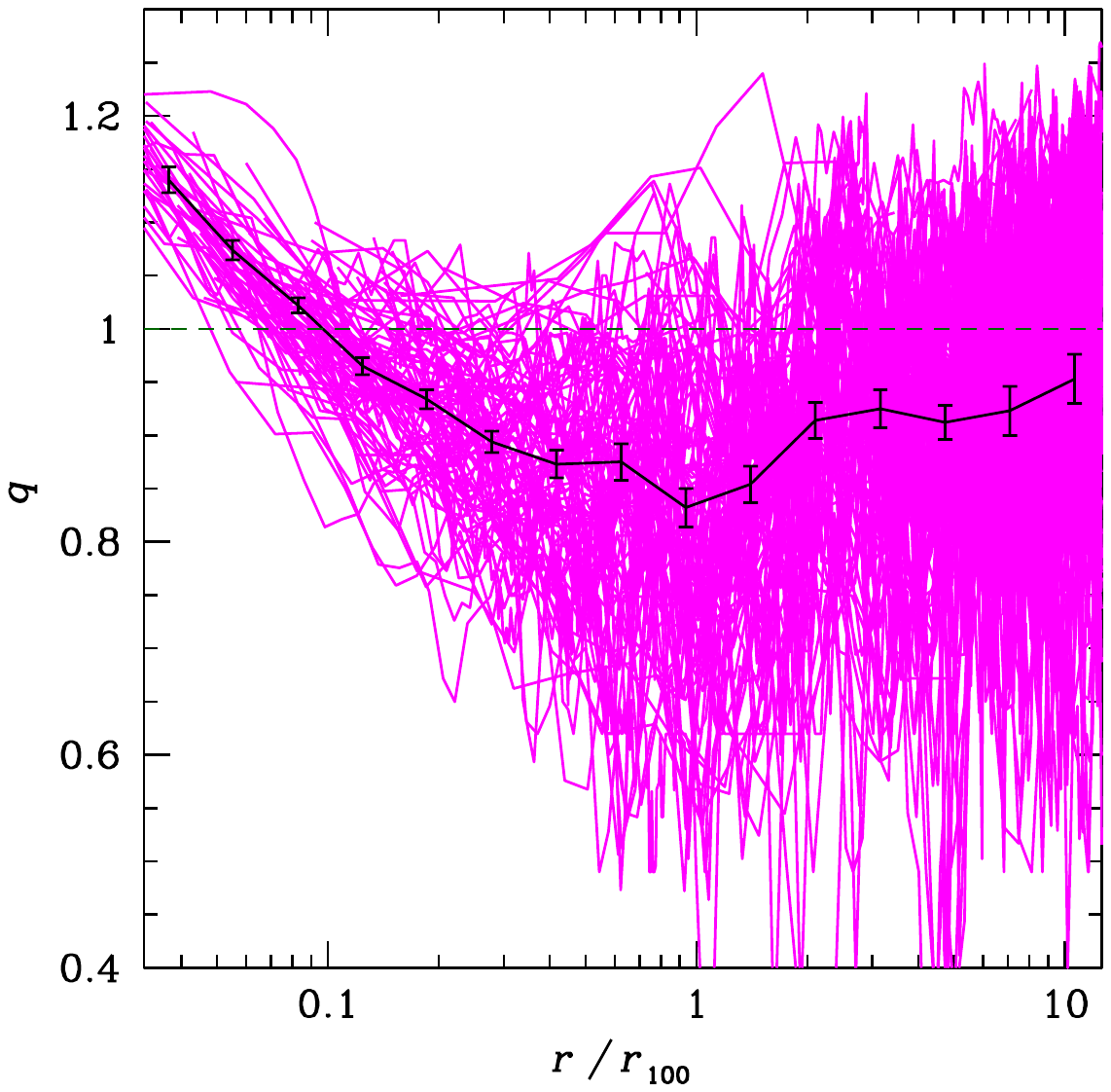} 
\caption{Non-Gaussianity index $q$ (best-fit)
  as a
  function of distance from halo centre (in virial
  units), 
for the 90 individual
  haloes (\emph{magenta lines}). 
The values of $q$ are obtained with maximum-likelihood
  (equation~\ref{eq_like2}) fits of
  equation~(\ref{eq_anisotropic_G}) to the distribution of
the dimensionless normalized speed $u$ (equations~\ref{eq_u_r}, \ref{eq_u_t},
and \ref{eq_u}).
The \emph{black line and error bars} are the medians and
  their uncertainties ($1.25\, \sigma/\sqrt{N}$, where 
$N$ and $\sigma$ are the number and standard deviation of $q$ values
  available for the given radial distance).
The \emph{dark green dashed horizontal line} shows the anisotropic Gaussian joint
velocity distribution ($q=1$).
} 
\label{img_qofrall}
\end{figure}
Fig.~\ref{img_qofrall} shows the best-fit $q$ of our anisotropic model
(equation~\ref{eq_anisotropic_G}) versus distance from the
halo centre for the 90 individual haloes. 
In this work, all fits of $q$ to the distribution of $u$  are performed by maximum likelihood estimation.
No halo exhibits a
Gaussian\footnote{By Gaussian, we refer to the $q=1$ limit of our anisotropic
  $q$-Gaussian model, not to be confused with the isotropic Gaussian of Sect.~\ref{sec_maxwell}.}
behaviour at all radii.
The non-Gaussianity index starts
above unity, decreases to unity at typically $r_{100}/10$, keeps decreasing to
$q \simeq 0.85$ at $r \approx r_{100}$, then rises rapidly to $q\simeq 0.94$
at $2-2.5\,r_{100}$, where it reaches a plateau. 

\begin{figure}
\includegraphics[width=\hsize,bb=0 140 600 700]{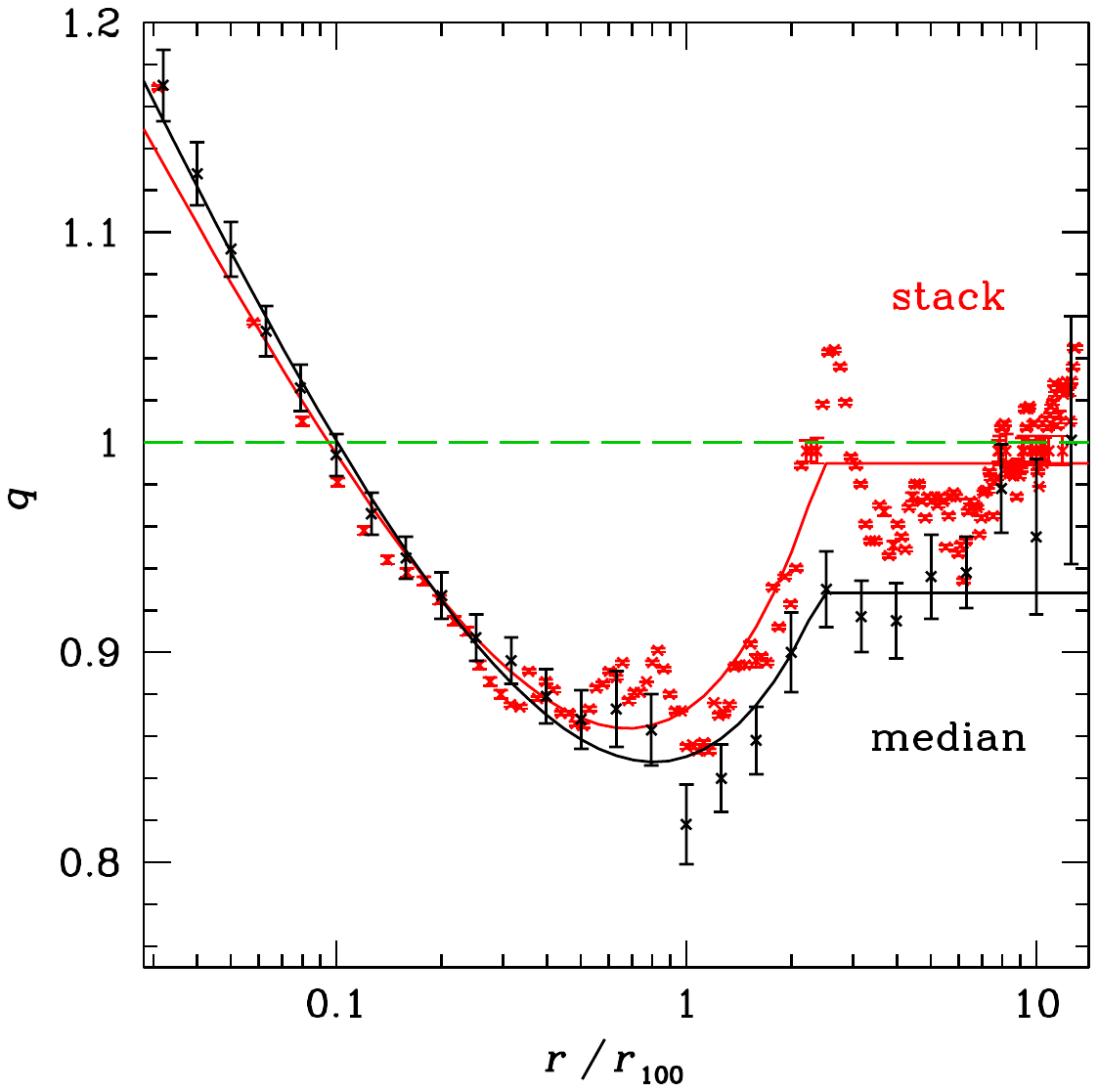} 
\caption{Non-gaussianity index $q$ (best fit) versus distance
  to the centre in virial units, for the median of the
$q(r)$ profiles (\emph{black}) and for the stacked halo
  (\emph{red}).
The curves are fits to $q(r)$
  using the model of equation~(\ref{eq_q_r2}). 
The uncertainties for $q$ of the stacked halo are from the fits, while those
for the median case the uncertainty on the median, measured as 
$1.25/\sqrt{90}$ times the standard
deviation of the 90 values.
}
\label{img_q_bestfit_stack_and_median_all}
\end{figure}

Fig.~\ref{img_q_bestfit_stack_and_median_all} shows the best fit values of $q$
obtained in the stacked halo as well as the 
median $q(r)$ of
individual haloes (see Sect.~\ref{sec_simulations}).
We can see how $q$ changes in
comparison to the Gaussian case of $q=1$ (plotted as a dashed horizontal line). We
note that $q(r) \propto - \log{(r/r_{100})}$ in the inner region, while  it rises approximately  as some power of $(r/r_{100})$ at
larger radii, until it
reaches a plateau near unity.

This behaviour of $q(r)$ can be  described with the following 5-parameter analytical function:
\begin{eqnarray}
q(r) &=& q_{\rm low} - a \,
\left(
{1-y^b\over b\,\ln 10}+\log y \right) \ , 
\label{eq_q_r1}\\
y &=& {{\rm Min}(r/r_{100},x_{\rm flat})\over x_{\rm low}}\ .
\label{eq_q_r2}
\end{eqnarray}
In equation~(\ref{eq_q_r1}), $a$ is the limit of ${\rm d}q/{\rm d}\log
r/r_{100}$ when $r\to 0$, $b$ is
close to the
power of $r/r_{100}$ in the rising portion of $q(r)$, while in
equation~(\ref{eq_q_r2}) $x_{\rm low}=r_{\rm
  low}/r_{100}$ is where $q(r)$
is minimized at $q(r_{\rm low}) = q_{\rm low}$, and $x_{\rm flat}=r_{\rm
  flat}/r_{100}$, such that 
$q(r)$ reaches its plateau at $r_{\rm flat}$.


The continuous lines in Fig.~\ref{img_q_bestfit_stack_and_median_all} show the
result of this fit. Table~\ref{tab_params_q_r} shows the values of the
parameters obtained in the ``stack''
and ``median'' cases.

\begin{table}
\begin{center}
\caption{Parameters of the best-fit $q(r)$ function of
  equations~(\ref{eq_q_r1}) and (\ref{eq_q_r2}) to the data of the 90 simulated haloes
\label{tab_params_q_r}}
\begin{tabular}{lccccc}
\hline
Method & $a$ & $b$ & $x_{\rm low}$ & $q_{\rm low}$ & $x_{\rm flat}$\\ 
\hline
Stack  & 0.331 & 0.757 & 0.683 & 0.864 & 2.45 \\
Median & 0.384 & 0.620 & 0.807 & 0.848 & 2.44 \\
\hline
\end{tabular}
\end{center}
\end{table}


\begin{figure}
\centering
\includegraphics[width=\hsize,bb=0 140 600 700]{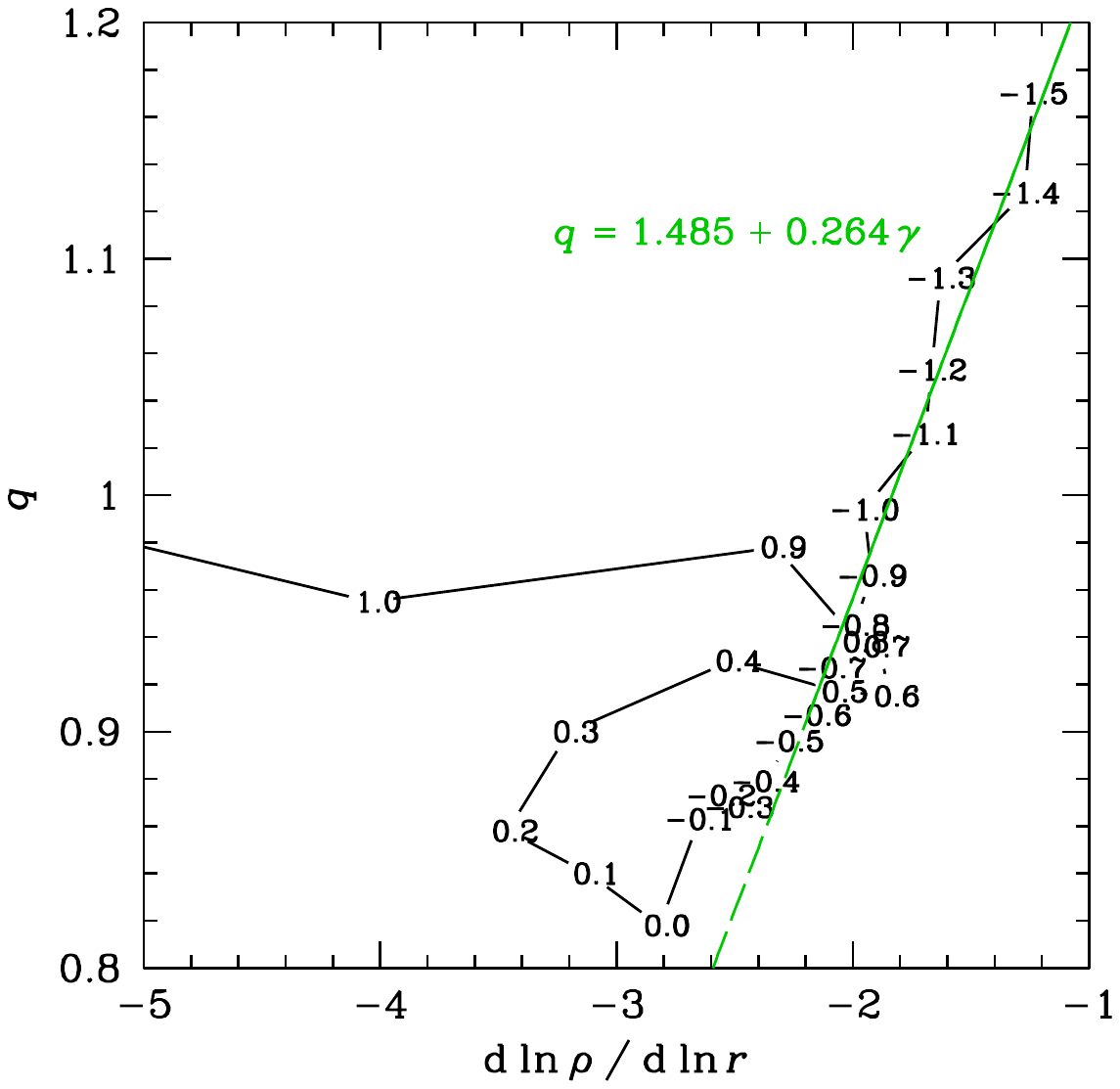} 
\includegraphics[width=\hsize,bb=0 140 600 700]{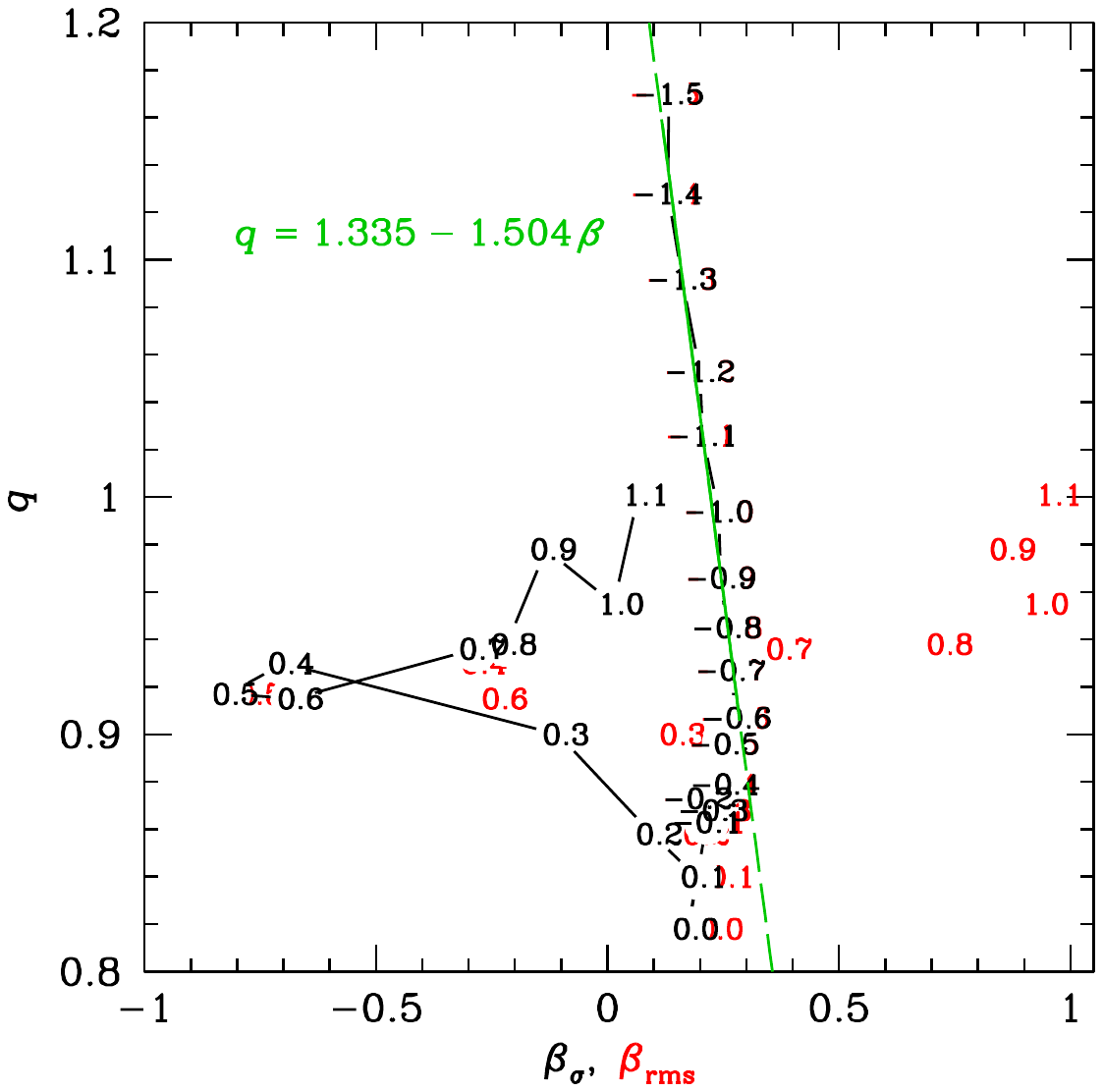} 
\caption{Relation between non-gaussianity index $q$ and the slope of the
  density profile (\emph{top})
and the velocity
  anisotropy (\emph{bottom}), defined using velocity dispersions
  (equation~\ref{betadef}, \emph{black} or using rms velocities, i.e. adding in quadrature
  mean streaming motions to the velocity dispersions, \emph{red}). All
  quantities refer to the medians of the 90 haloes. The points are labelled by their value of
  $\log_{10}(r/r_{100})$. The green lines show linear fits in the region
  $r/r_{100}<0.3$ (shown with solid type), with the coefficients labelled (with $\gamma={\rm
    d}\ln\rho/{\rm d}\ln r$).
}
\label{Hansenplots}
\end{figure}
Interestingly, as shown in Fig.~\ref{Hansenplots}, the non-Gaussianity
index $q$ of our 2D velocity model (equation~\ref{eq_anisotropic_F_ur_ut}) 
is linearly related to both the logarithmic slope of the density profile,
$\gamma={\rm d}\ln\rho/{\rm d}\ln r$, and to the  velocity anisotropy
$\beta$.
This agreement is quasi-perfect for radii  $r/r_{100}<0.4$ ($\log
r/r_{100}<-0.4$),  very good for $r/r_{100}\leq 0.5$ ($\log
r/r_{100}\leq -0.3$), and decent up to the virial radius.
Figure~\ref{Hansenplots} is for the median properties of the 90 haloes. For
the stack of the 90 haloes, the analogous figure is similar, but with a faster
divergence from the linear relation starting at $\log r/r_{100} = -0.3$.

The linear relation between $q$ and $\gamma$ confirms the linear trend of
non-Gaussianity with density slope previously suggested by \cite{HMZS06} for
the radial and tangential components of the velocity distribution, although
here the linearity is confined to $r < 0.4\,r_{100}$ (for a precise
linearity, and up to the virial radius for approximate linearity).

We also find linear trends of $q$ with velocity anisotropy, using  two different
measures of velocity anisotropy, the one
(hereafter $\beta_\sigma$)
using velocity dispersions (equation~\ref{betadef}),  and the one using rms
velocities, $\beta_{\rm rms} = 1 - (\langle v_\theta^2\rangle +\langle
v_\phi^2 \rangle )/(2 \langle v_r^2\rangle)$. 
Note that both measures of anisotropy,
$\beta_\sigma$ and $\beta_{\rm rms}$,
diverge from one another 
only beyond the virial radius, where radial streaming motions become important.

\subsection{Goodness-of-fit}
We now ask whether  the best-fit $q$-Gaussian model provides an adequate
representation of the simulated data. 
\begin{figure}
\includegraphics[width=\hsize]{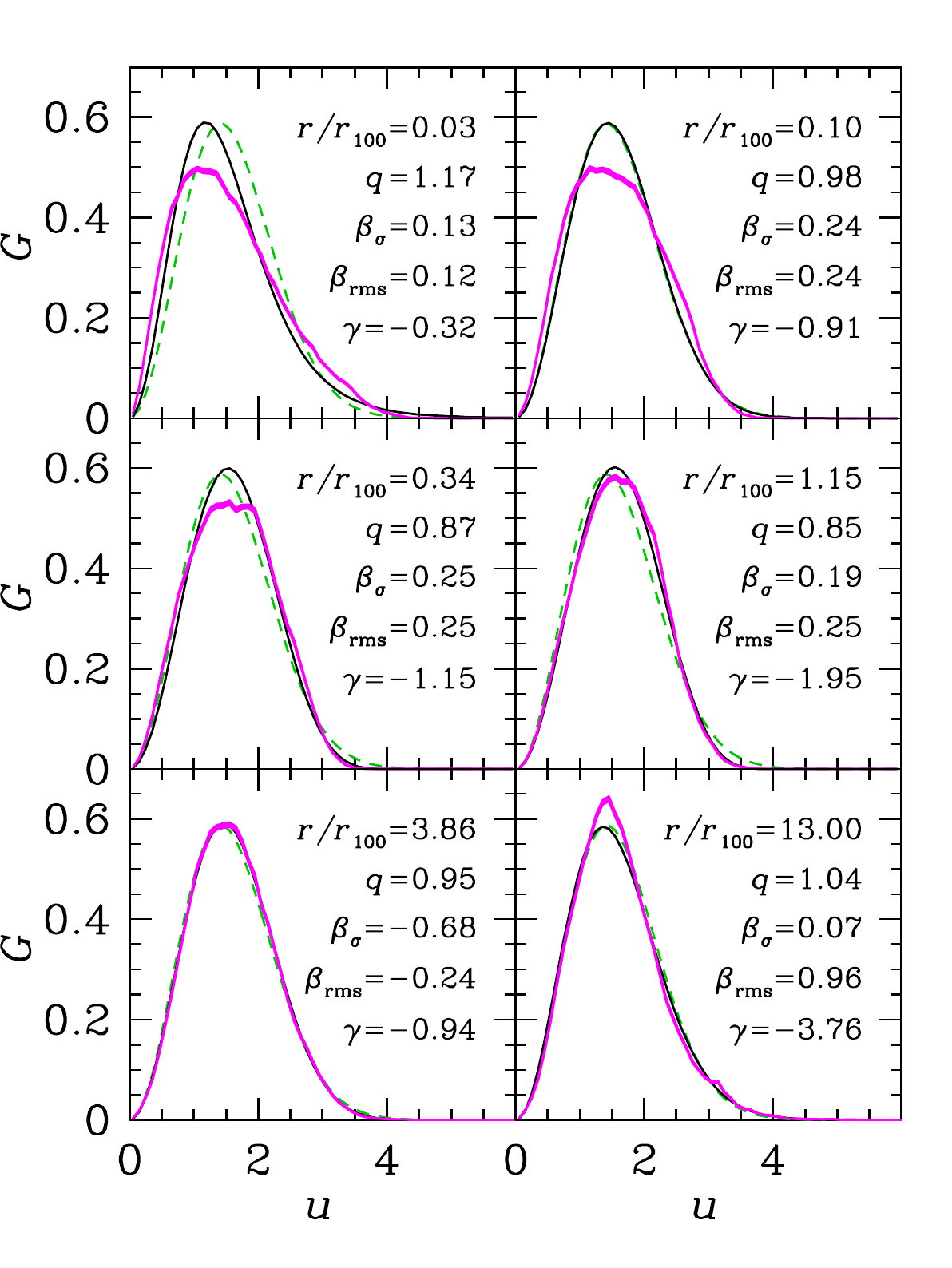}
\caption{Speed distribution function (equations~\ref{eq_u_r}, \ref{eq_u_t},
and \ref{eq_u}), at six different
  radii of the stacked halo (300$\,$000 particles per radial bin). 
The \emph{thin magenta shaded area} represents the data with the uncertainties
  calculated with bootstraps. The \emph{continuous black curve} represents
  the prediction of the anisotropic $q$-gaussian
  (equation~\ref{eq_anisotropic_G}), and the \emph{dashed green line} represents
  the  (anisotropic) Gaussian
  prediction. The radii, best-fit values for $q$,  velocity anisotropies
  measured using velocity dispersions ($\beta_\sigma$) or rms velocities
  (i.e., including streaming motions, $\beta_{\rm rms}$) and logarithmic
  slopes of the density profile ($\gamma$) are shown 
in 
each panel. 
}
\label{img_G_u_some}
\end{figure}
Fig.~\ref{img_G_u_some} displays the distribution of dimensionless normalized 
SDFs measured in the simulated haloes in the stacked case
as well as the best fit Gaussian and $q$-Gaussian SDFs. 
The dimensionless normalized SDF reaches its mode at low normalized velocity modulus at low radii ($r
\leq 0.34\, r_{100}$) at ``normal'' values at the virial radius and $4\,
r_{100}$, and at super-normal values at very high radii ($13\,r_{100}$). 

We see that within $r \la 0.34\,r_{100}$ (upper and middle-left panels of Fig.~\ref{img_G_u_some}),
the $q$-Gaussian does not describe well the data (nor does the Gaussian,
since it is a special case of the $q$-Gaussian).
For example, at $r = 0.03\, r_{100}$, while the $q$-Gaussian with $q=1.17$ fits the data better
than the Gaussian (mostly in the low-end tail of $G(u)$), 
its predicted velocity distribution  is too peaked, and
it presents an important excess of very high ($u>4$) velocities.
These two characteristics
are also
present at $r = 0.1\,r_{100}$.
On the other hand, the non-Gaussian model fits well the SDF of the simulations
near the virial radius, while the Gaussian model fits much less well.
At 4 virial radii, both the Gaussian and non-Gaussian models fit well the
simulated SDF. 

Interestingly, the flattening of the SDF near its mode (i.e. flat top SDF) appears to be related to
radius and slope of the density profile, as it is most prominent at small
radii and shallow density slopes, while it appears unrelated to the
velocity anisotropy.

A quantitative measure of the goodness-of-fit can be obtained by the
Kolmogorov-Smirnov (KS) test of the maximum absolute difference between the
cumulative distribution functions (CDFs) of the predicted and the simulated
dimensionless normalized speeds.
Given the large number of points in each radial bin of the stacked sample
($\approx 300\,000$), the probability that the
$q$-Gaussian model is an adequate representation of the data is rejected at
over 95\% confidence if the maximum absolute difference between the
cumulative distribution functions of the model and the data is above 0.0025.

As we shall see below, 
the KS test rejects both Gaussian
and non-Gaussian models, at virtually all radii for the stacked halo and at
all radii when considering the median of the individual haloes.

Nevertheless, given that we have roughly equal numbers of particles
per radial bin in both cases, we can illustrate the result of the KS test by
plotting the maximum absolute difference in the CDFs. 
\begin{figure}
\includegraphics[width=\hsize,bb=0 140 600 700]{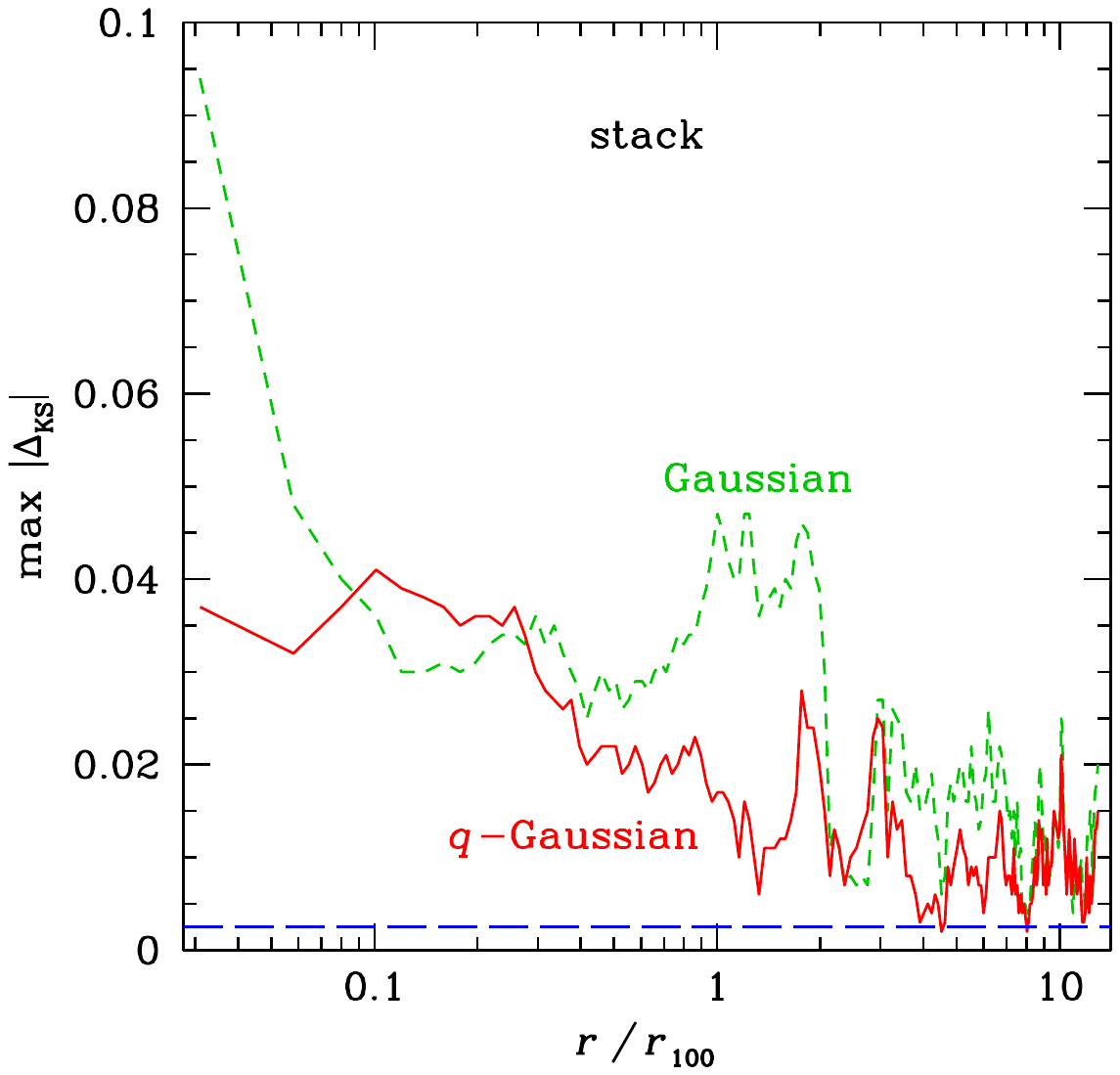}
\caption{Maximum absolute difference between the cumulative distribution functions
  of $u$ (see equation~\ref{eq_anisotropic_G}) of the stacked halo and those for
  the best-fit 
Gaussian (\emph{dashed green line}) and $q$-Gaussian (\emph{continuous red line}) models. 
The \emph{blue dashed horizontal line} represents the upper limit for a model
that cannot be rejected with greater than 95 percent confidence.
}
\label{img_Delta_KS_stack}
\end{figure}
Figure~\ref{img_Delta_KS_stack} shows that when the KS test is applied to the
stacked halo, 
the $q$-Gaussian model leads to a  better representation of the
distribution of $u$ than does the Gaussian model, at small radii ($r <
0.06\,r_{100}$) and near the virial radius.
Of course, given its extra parameter, one expects the $q$-Gaussian to perform
at least as well as the Gaussian (it can occasionally perform slightly worse,
because the KS test does not measure the agreement between model and data in
the same way as maximum likelihood estimation.
As mentioned above, the non-Gaussian model can nevertheless be rejected with
95 percent confidence, 
except for two radii (out of 175, where the red line passes below the blue
horizontal dashed line).

\begin{figure}
\includegraphics[width=\hsize,bb=0 140 600 700]{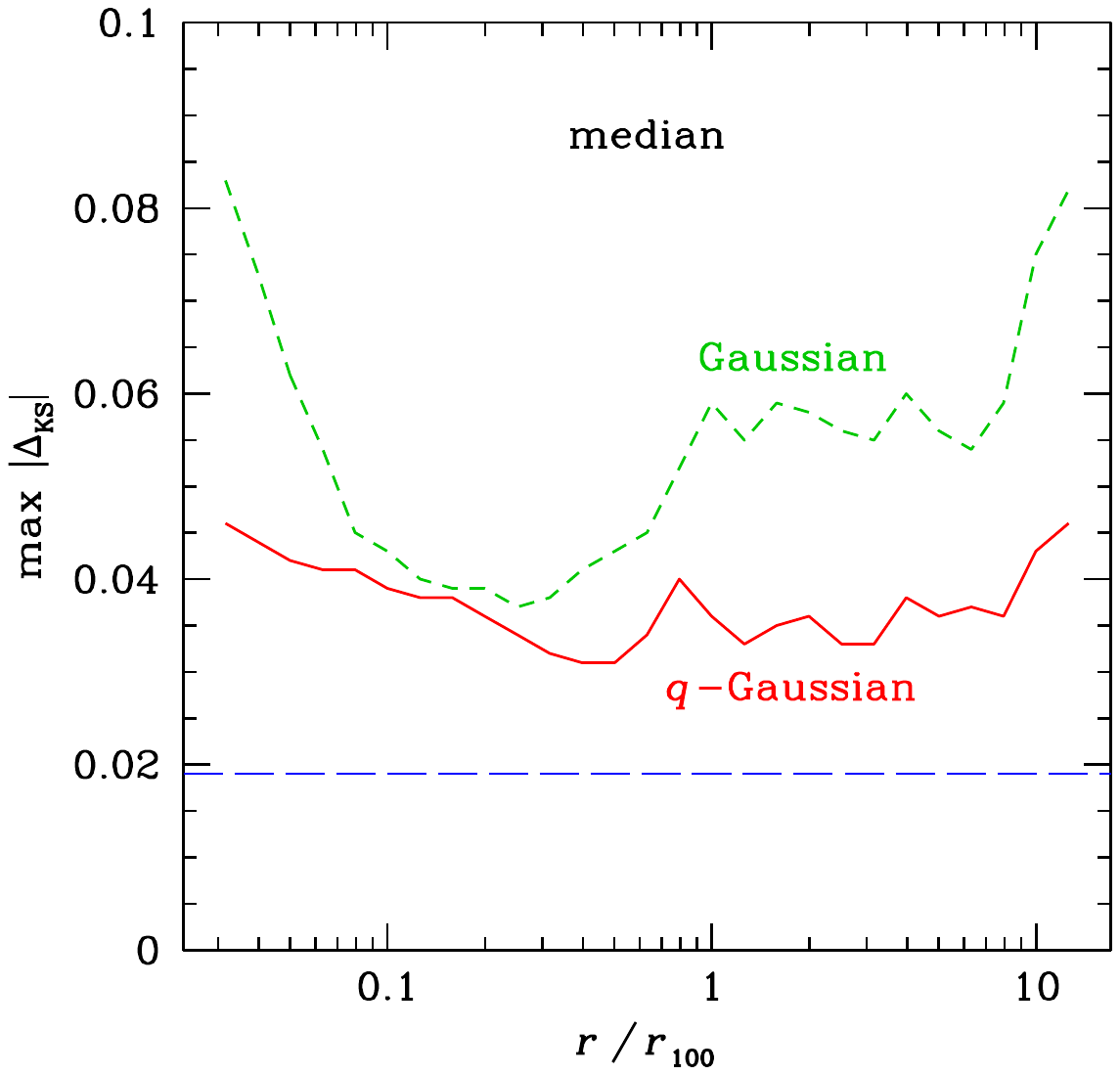}
\caption{Median of the 90 maximal absolute differences between the cumulative
  distribution functions of $u$ (see equation~\ref{eq_anisotropic_G}) of the
  individual haloes and those for the
  best-fitting $q$-Gaussian   (\emph{continuous red line}) or Gaussian
  (\emph{dashed green line})  models.
The \emph{blue dashed horizontal line} represents the upper limit for a model
that cannot be rejected with greater than 95 percent confidence.
}
\label{img_Delta_KS_median}
\end{figure}
%
Fig.~\ref{img_Delta_KS_median} shows the median KS test results over 90
haloes, i.e. the median value of the 90 maximal absolute differences in the CDFs. One sees that
the $q$-Gaussian fits are typically better not only at small radii ($r <
0.05\,r_{100}$), and near the virial radius, but also at all radii above the
virial radius. Nevertheless, as mentioned above, the $q$-Gaussian model
is typically rejected (from the median value of max $\Delta_{\rm KS}$) with
over 95 percent confidence at all radii. 

We will check in Sect.~\ref{sec_bayes} below whether the smaller discrepancies for the
$q$-Gaussian model (with the simulated data) 
in comparison with those of the Gaussian CDF are statistically significant, given the extra parameter of the non-Gaussian model.

\begin{figure}
\includegraphics[width=\hsize]{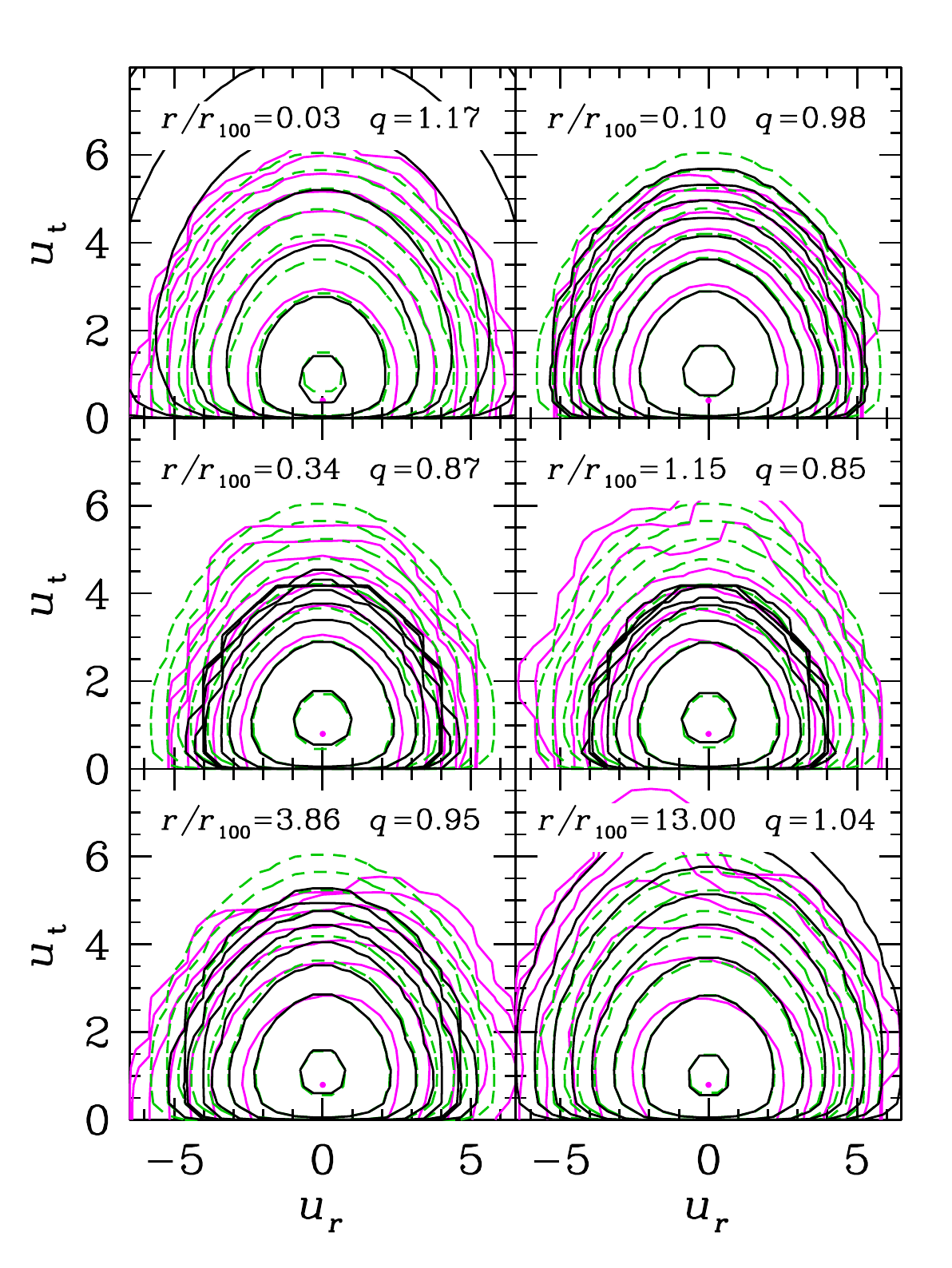}
\caption{Contour plots of the VDF $F(u_r,u_{\rm t})$ as a function of 
the radial and tangential components of the
dimensionless normalized velocities
of the stacked halo, for the same six radial bins as in
  Fig.~\ref{img_G_u_some}. 
The colors and radii in each plot are the same as in Fig.~\ref{img_G_u_some}.
The contours are
  logarithmically-spaced ($\approx 0.6$ dex), and those of the models follow the same levels as
  those of the data (the outermost  $q$-Gaussian (\emph{black}) contour does
  not appear for $r/r_{100} = 0.03$).
}
\label{img_G_contours_some}
\end{figure}
Another way to qualitatively evaluate the goodness-of-fit is to observe the
contours for the 2-dimensional velocity distribution defined by the radial and
tangential components.
This is shown in Fig.~\ref{img_G_contours_some}.
While the basic shapes of the contours of the models match fairly well the
contours of the simulated data, there are differences, in the shapes, and,
more strikingly, in the extent of the contours.

At $r = 0.03\, r_{100}$, the contours of the best-fit $q$-Gaussian extend to
much greater combinations of $u_r$ and $u_{\rm t}$, i.e. to greater
dimensionless normalized speeds. This is another sign that the $q$-Gaussian predicts
much more very high velocity objects than is seen in the simulation.
%
The better fit of the $q=1.17$ $q$-Gaussian in
comparison with the Gaussian is caused by the difference in the low-end tail
($u<0.8$) of $G(u)$ for these two cases (see upper left
panel of Fig.~\ref{img_G_u_some}), which is difficult to distinguish in the
contours.
At $r=0.1\,r_{100}$, the best-fit value of $q$ is near unity,
so the contours of the Gaussian and $q$-Gaussian
are identical. Moreover, they are quite similar to the contours extracted from
the simulated data (magenta), indicating that the high-end of the VDFs are similar (see
upper-right panel of Fig.~\ref{img_G_u_some}).
 At $r=0.34$ and $1.15\,r_{100}$, the data contours are more extended in
both $u_r$ and $u_{\rm t}$ than the best-fit $q$-Gaussian. This occurs at
very high speeds ($u > 5$) and is difficult to see in the SDF (Fig.~\ref{img_G_u_some}).
At $r=3.9\,r_{100}$, the contours of the best-fit $q$-Gaussian model matches
those of the data for
the tangential velocities, but the data contours extend to greater absolute radial velocities.
Finally, at $r=13\,r_{100}$, the contours of the best-fit $q$-Gaussian model
match fairly well the data contours.

At all radii, the most probable pair of $(u_r,u_{\rm t})$ is at lower
tangential velocity than predicted by the $q$-Gaussian and Gaussian models,
and low tangential velocities at high absolute radial velocities are not
avoided contrary to the model predictions (near $u_{\rm t}=0$, the model contours
move inwards while the data contours do not).


\subsection{Does the $q$-Gaussian model reproduce the data  significantly
  better than the Gaussian?}
\label{sec_bayes}
We now ask whether the $q$-Gaussian provides a significantly better fit to
the simulation data than does the Gaussian, taking into account the extra
parameter involved in the former.
For this, we evaluated two measures of \emph{Bayesian
  evidence},\textcolor{red}{\footnote{\textcolor{red}{In the published version, equations~(\ref{AIC}) and (\ref{BIC}) contained
  typographic errors, with erroneous `ln' terms in front of $N_{\rm pars}$.}}}
 the
\emph{Akaike Information Criterion} (AIC, \citealp{Akaike73}), 
\begin{equation}
\textcolor{red}{{\rm AIC} = -2\ln {\cal L}_{\rm max} + 2\, N_{\rm pars}} \ ,
\label{AIC}
\end{equation}
corrected for finite sample size by
\cite{HurvichTsai89}, to yield the 
\emph{corrected Akaike Information Criterion} (AICc) as
\begin{equation}
{\rm AICc} = {\rm AIC} + 2 \,{N_{\rm pars}(N_{\rm  pars}+1)\over N_{\rm
  data}-N_{\rm pars}-1} \ ,
\label{AICc}
\end{equation}
 as well as the
\emph{Bayes Information Criterion} (BIC, \citealp{Schwarz78}),
\begin{equation}
\textcolor{red}{{\rm BIC} =  -2\ln {\cal L}_{\rm max} + \ln N_{\rm data}\,N_{\rm pars}}
\ .
\label{BIC}
\end{equation}
For AIC, AICc and BIC, the probability that one model is better than another
is $P = \exp(-\Delta{\rm IC})/2$, where $\Delta$IC is the difference between
two fits of any of the three information
criteria \citep{KassRafferty95}. Thus, one can conclude that one model is superior to the other with
95\% confidence if its value of IC is $2\ln(0.05) \simeq -5.99$ greater than
the IC of the other model.
Comparing equations~(\ref{AIC}) and (\ref{BIC}), one easily sees that BIC
penalizes more the extra parameter(s) in the presence of large data sets.
The choice between AICc (or AIC) and BIC is still debated \citep{BurnhamAnderson04,
  Trotta_2008}. However, for our purposes, the results are extremely similar, so we
will only display BIC for clarity.
\begin{figure}
\centering
\includegraphics[width=\hsize,bb=0 140 600 700]{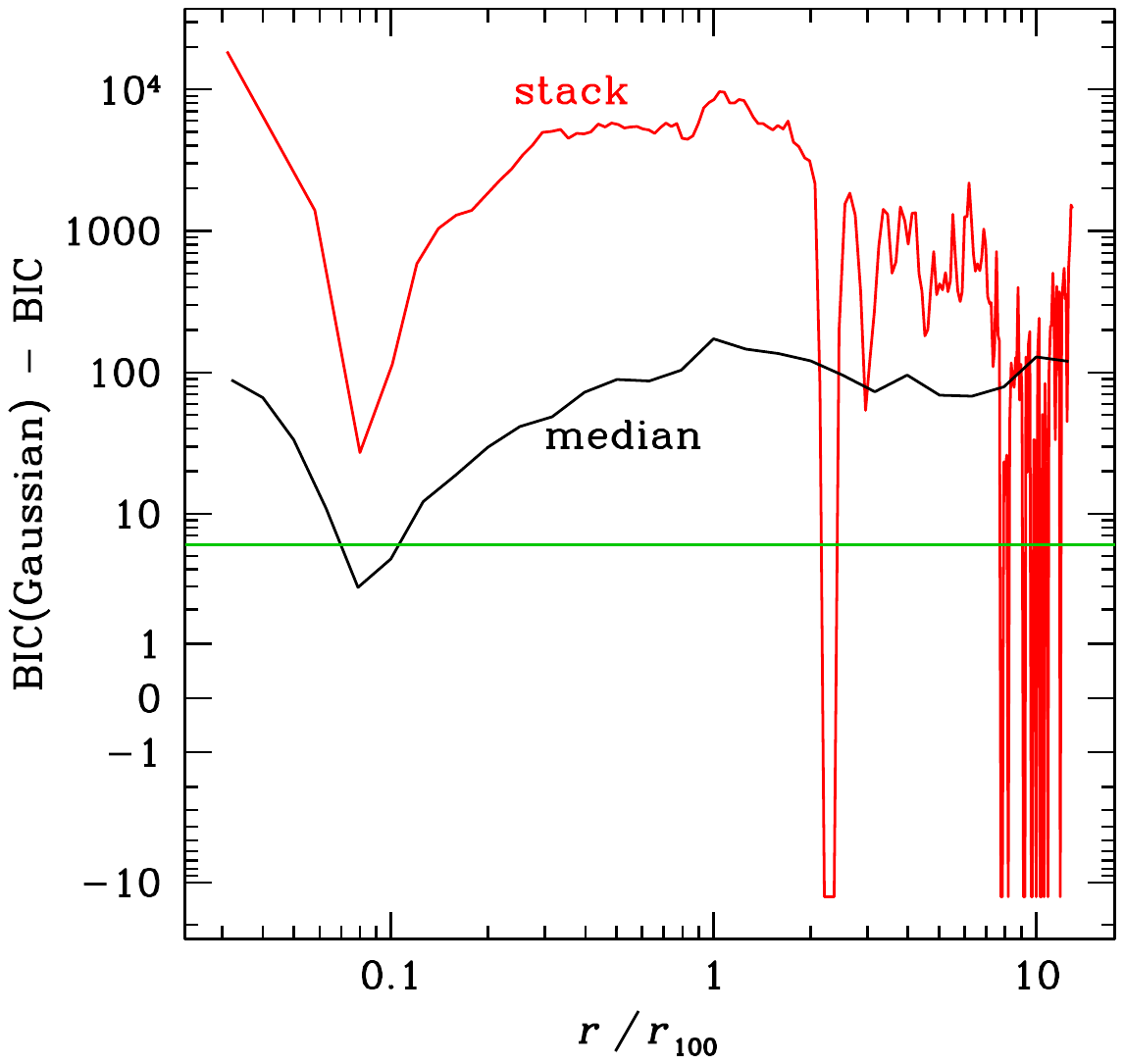} 
\caption{Bayes Information Criterion 
(Eq.~[\ref{BIC}])
obtained by
  fitting the non-Gaussianity index $q$ to the stacked halo (\emph{red
    broken line}) and to the  median of halos (\emph{continuous black smoother
    line}).
The $y$ axis follows an arcsinh scaling.
\textcolor{red}{Values above the \emph{horizontal green line} indicate that the $q$-Gaussian
model is a better fit then the Gaussian one, even when considering its extra parameter.}
For the stack, there is strong evidence that the $q$-Gaussian model is a
better representation than the Gaussian at all radii except $r=0.1$ and 1.5
virial radii. For the median of the fits, the evidence that the $q$-Gaussian is a
better representation than the Gaussian is strong at all radii, but the
coarse grid is missing $r\approx 0.1 r_{100}$, where $q\simeq 1$, hence there
should be negative evidence (as in the case of the stacked halo).}
\label{img_BIC}
\end{figure}
%


Fig.~\ref{img_BIC} shows that there is strong evidence that the $q$-Gaussian is preferable
to the Gaussian distribution for the stacked halo (red line), except, of
course, at the points where $q \approx 1$, where the 
addition of the index $q$ is not necessary. For the median of the fits to
the 90 haloes
(smoother black line), we have strong evidence in favor of the $q$-Gaussian at
all radii. However, with a finer grid we would necessarily find no evidence
in favor of the $q$-Gaussian at $r \simeq 0.1\,r_{100}$, where $q\simeq 1$.


\subsection{Mass dependence}

We now investigate a possible dependence of the index $q$ on the mass of
the haloes. To do this, we divide our sample into the 3 mass subsamples of 30
haloes each (see Sect.~\ref{sec_simulations}). 
\begin{figure}
\includegraphics[width=\hsize,bb=0 140 600 700]{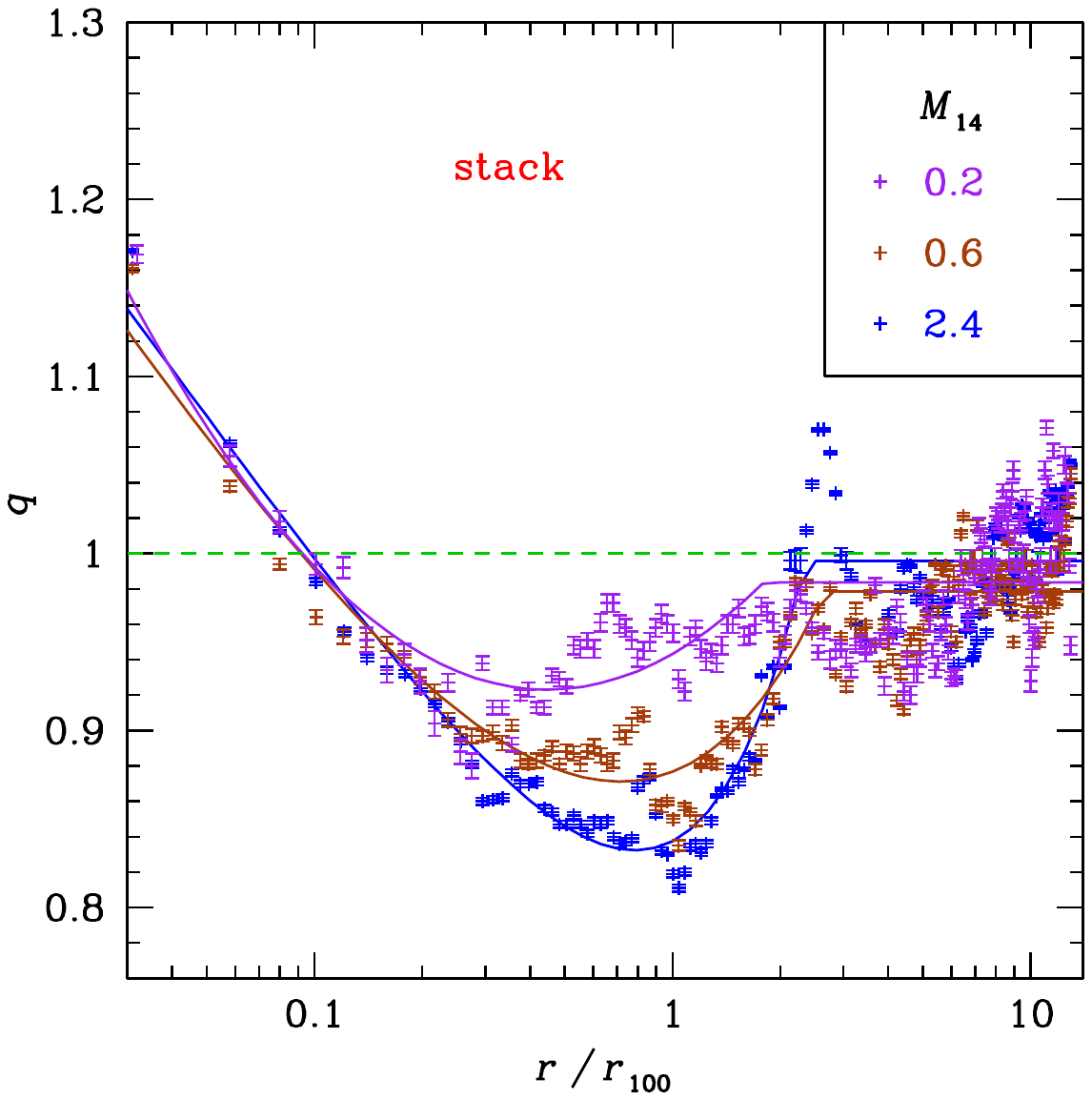}
\caption{Best-fitting value of $q$ for the stacked halo for 3 mass subsamples (\emph{purple}, \emph{brown}
  and \emph{blue} in increasing order of mass (the values of $M_{14} =
  \langle M \rangle_{\rm bin}/10^{14} {\rm M}_\odot$ are shown and are in
  decreasing order of $q$ at $r=r_{100}$). The \emph{green horizontal
    dashed line} is the Gaussian.)}
\label{img_q_bestfit_stack}
%
\includegraphics[width=\hsize,bb=0 140 600 700]{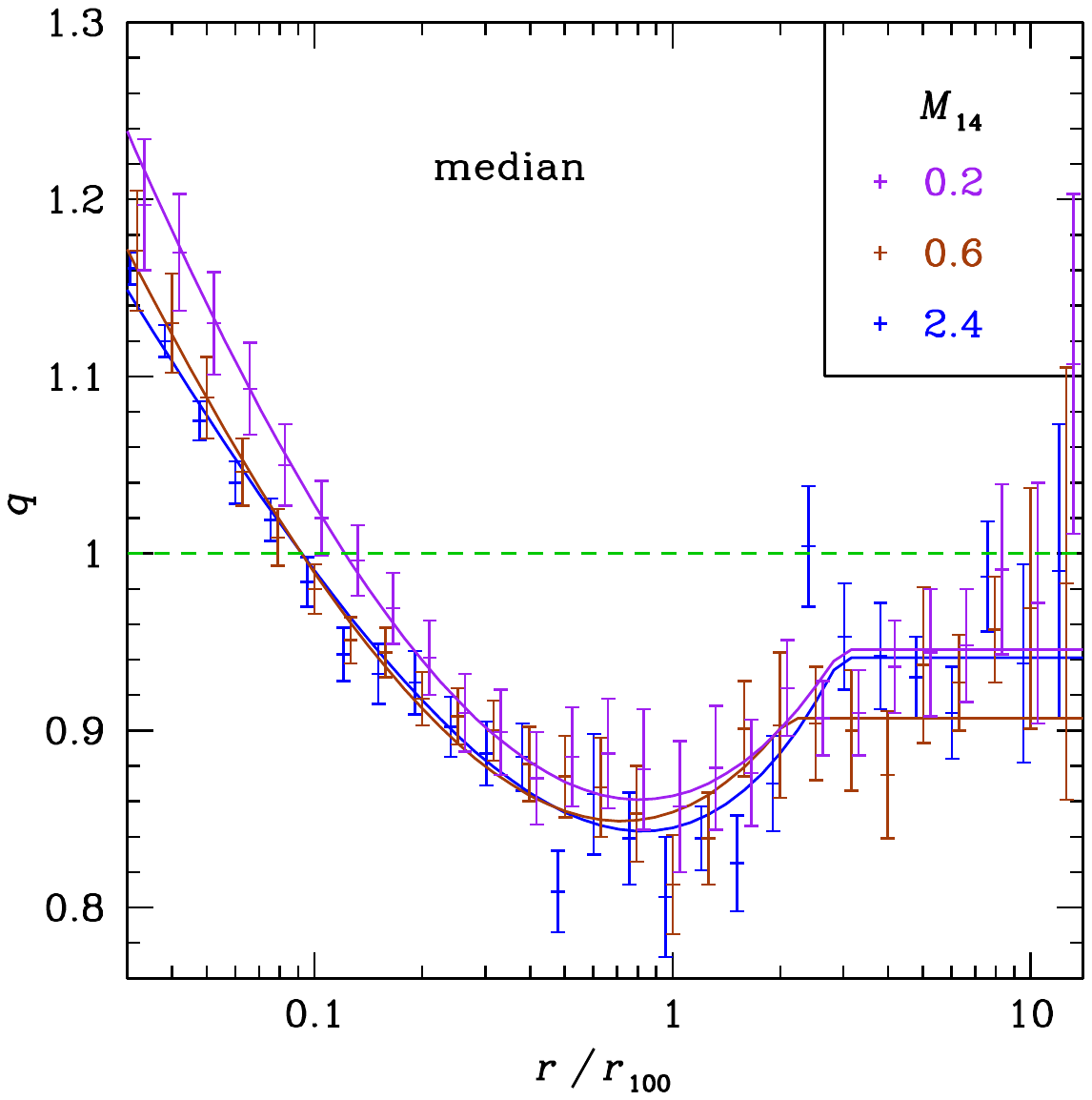}
\caption{Same as Fig.~\ref{img_q_bestfit_stack}, but for the  median of the
  30 $q(r)$ profiles in each subsample}
\label{img_q_bestfit_median}
\end{figure}

For each of these 3 subsamples, we performed the same fit procedure as
before, for the stacked halo and the individual haloes, to which we consider
the median fit.
Fig.~\ref{img_q_bestfit_stack} shows that the $q(r)$ profiles of the 3 stacked
haloes show reasonable differences at intermediate radii ($0.3 < r/r_{100} < 2$):
The minimum $q$ is lower for the highest mass bins and this minimum is
reached at progressively larger radii (in units of the virial radius).
This is confirmed by the fits of the analytical function of
equations~(\ref{eq_q_r1}) and (\ref{eq_q_r2})
to the measured $q(r)$, as provided in Table~\ref{tab_params_q_r_stack}.

On the other hand, Fig.~\ref{img_q_bestfit_median} indicates that the 3 subsamples produce
very similar median $q(r)$ profiles (see the parameters of the analytical
function fit to the median $q(r)$ listed in
Table~\ref{tab_params_q_r_median}). The minimum $q$ is again lower for
increasingly higher halo masses, but this modulation is much weaker than for
the stacked haloes: the analytical fits indicate that the differences in
$q_{\rm low}$ between the highest and lowest mass bins is only 0.018 for the
median fits in comparison with 0.091 for the fits to the 3 stacked haloes.

\begin{table}
\begin{center}
\caption{Parameters of the best-fit $q(r)$ function of
  equations~(\ref{eq_q_r1}) and (\ref{eq_q_r2}) to the stacked data split in 3 bins of halo mass
\label{tab_params_q_r_stack}
}
\begin{tabular}{cccccc}
\hline
$\langle M\rangle/{\rm M}_\odot$ & $a$ & $b$ & $x_{\rm low}$ & $q_{\rm low}$ & $x_{\rm flat}$\\ \hline
$1.55\times 10^{14} $ &  \ \,0.279 &      1.329 & 0.787 &       0.832 &       2.33 \\
$6.10\times 10^{13} $ &  \ \,0.318 &     0.662 & 0.715 &      0.871 &      2.65 \\
$1.99\times 10^{13} $ &  20.257 &      0.0071 &     0.446 &      0.923 &      1.80 \\
\hline
\end{tabular}
\end{center}
Notes: Values are for the 3 mass subsamples, whose median is indicated in the first column.
\end{table}

\begin{table}
\begin{center}
\caption{Parameters of the median best-fit $q(r)$ function of
  equations~(\ref{eq_q_r1}) and (\ref{eq_q_r2}) to the median data of individual haloes, split
  in 3 bins of halo mass
\label{tab_params_q_r_median}
}
\begin{tabular}{cccccc}
\hline
$\langle M\rangle/{\rm M}_\odot$ & $a$ & $b$ & $x_{\rm low}$ & $q_{\rm low}$ & $x_{\rm flat}$\\ \hline
$1.55\times 10^{14} $ &  0.393 &     0.553 & 0.822 &      0.843 &      2.93 \\
$6.10\times 10^{13} $ &  0.597 &     0.344 & 0.720 &      0.849 &      2.10 \\
$1.99\times 10^{13} $ &  0.813 &     0.256 & 0.810 &      0.861 &      2.95 \\
\hline
\end{tabular}
\end{center}
Notes: Values are for the 3 mass subsamples, whose median is indicated in the first column.
\end{table}

\section{Conclusions and Discussion}
\label{sec_discuss}
In this work, we propose, for the first time, a model of the velocity
distribution function  for $\Lambda$CDM haloes  that is non separable in its radial and tangential
components and only involves a single free parameter: 
it is an anisotropic version of the $q$-Gaussian velocity
distribution, given by equation~(\ref{eq_anisotropic_F_general}), or equivalently
equations~(\ref{eq_anisotropic_F_u}), (\ref{eq_anisotropic_F_ur_ut}) or
(\ref{eq_anisotropic_G}), which is built on
an isotropic dimensionless normalized
velocity field $\bmath u$.
Our VDF involves a single dimensionless normalized velocity, the modulus of
$\bmath u$, which can be written (for 
negligible streaming motions expected with the virial radius)
\begin{equation}
u={\sqrt{v_r^2 + v_{\rm t}^2/( 1-\beta)}\over \sigma_r} \ ,
\label{eq_u_v}
\end{equation}
(see equations~\ref{betadef} and \ref{eq_u}), and a single value of non-Gaussianity. 
In other words, at a given radius $r$, the velocity distribution function is
a function of $\{2[E-\Phi(r)]+\beta(r) J^2/r^2\}/\sigma_r^2(r)$ instead of
$2[E-\Phi(r)]$ for the isotropic distribution.
We find
this parametrization preferable to one that is separable in the radial
and tangential coordinates, as the latter 
is inconsistent with the separable DF of $\Lambda$CDM haloes \citep{Wojtak+08},
although nothing guarantees that our anisotropic VDF does not violate 
 the 
Jeans theorem.
Our one-parameter VDF is also simpler to handle than the 3-parameter VDF of
\cite{Hunter14}, both of which are the only VDFs proposed so far involving
both radial and tangential velocities in an anisotropic fashion.

We test this anisotropic $q$-Gaussian VDF
by fitting the non-Gaussianity index $q$ at different radii of simulated haloes in
two ways: fitting after stacking all the selected haloes or fitting individual
haloes and calculating the median value of $q$ in each radial shell. 
We find that nearly all haloes show $q(r)$ decreasing with radius, from above
unity at very small radii to lower than unity near the virial radius. 
At low radii ($< 0.4\,r_{100}$), the
median $q$ is linearly related to both the
logarithmic slope of the density profile and the velocity anisotropy,
and this trend remains nearly linear up to the virial
radius
(Fig.~\ref{Hansenplots}).
Above the virial radius, both the median $q(r)$ and the $q(r)$ of the stacked halo
rise again to near unity and remain at that value up to the largest radii we
analyzed ($13\,r_{100}$), see Figs.~\ref{img_qofrall} and also
\ref{img_q_bestfit_stack_and_median_all}.

The anisotropic $q$-Gaussian predictions cannot match the simulated
data in a statistically significant way (Figs.~\ref{img_Delta_KS_stack} and
\ref{img_Delta_KS_median}).
The speed distribution functions of haloes have flatter tops at low
radii ($r \leq 0.34 \,r_{100}$) than the
model predictions, and cuspier tops at very high radii ($r=13\,r_{100}$), see
Fig.~\ref{img_G_u_some}. The models cannot produce too few very high speeds
at intermediate radii (0.34 and $1.1\,r_{100}$), but too many
high speeds at very low radii ($0.03\,r_{100}$), see Fig.~\ref{img_G_contours_some}.

Nevertheless, the anisotropic $q$-Gaussian is highly preferred to the Gaussian
distribution at nearly all radii, even when taking the extra parameter of the former into
consideration (Fig.~\ref{img_BIC}).

We provide appropriate expressions to describe the behaviour of the best
values of $q$ as a function of the distance to the centre (Tables 1--3). These expressions
can be used either to model individual haloes (the median case) or to model
stacked haloes. 

The linear relation of $q$ decreasing with increasing velocity
anisotropy parameter  $\beta$ (lower panel of Fig.~\ref{Hansenplots})
confirms the radial trends of $\beta$ and radial kurtosis  up to 2 virial radii seen in
fig.~3 of \cite{WLGM05}.
One may wonder whether the linear relation between $q$ and $\beta$ can be explained from first principles.
Our results are in line with the combination of 1)  $q$ increasing linearly
with decreasing density slope $\gamma$ found for
the radial VDF by \cite{HMZS06}  (at radii where the density profile has a slope shallower than $-2.5$,
roughly the virial radius) 
and 2) the wide-wing tangential VDF found by
\cite{HS12}, which could be assimilated to a $q$-Gaussian with constant
$q>1$. 
Our linear $q-\beta$ relation follows naturally
from the linear $\beta-\gamma$ relation at these radii \citep{HM06}, and our
linear $q-\gamma$ relation.

Still, the origin of the $q-\beta$ relation is not clear. Our VDF is
isotropic when expressed in terms of the dimensionless velocities obtained
after subtracting the mean velocities and dividing by the dispersions. So
our VDF ought to be independent of the velocity anisotropy.


The radial profile of $q(r)$ of the
stacked halo follows the median $q(r)$ profile at low radii, but
reaches 
somewhat greater values at large radii than the
median $q(r)$ of the 90 individual haloes: with asymptotic values near 0.99
for the stack versus 0.93 for the median of the individual fits 
(Fig.~\ref{img_q_bestfit_stack_and_median_all}).
This small difference may be due to a geometrical
effect produced in the stacking process. Indeed, haloes in $\Lambda$CDM
cosmological simulations are triaxial (e.g., \citealp{JS02}), but our
stacking was done without previously rotating the haloes to align their
principal axes. 
So the stacked $q(r)$ profile is estimated by analyzing a spherically
symmetric system, while the median $q(r)$ profile is the median of individual
triaxial systems (analyzed in concentric spherical shells).
Moreover, the velocity
ellipsoids of simulated haloes are aligned with their density ellipsoids
\citep{WGK13}, which should also affect the measure of $q$ in concentric
spherical shells.
However, the triaxiality of cosmologically simulated haloes is strongest in
the inner regions \citep*{JS02,SFC12}, and the alignment of the velocity ellipsoid with the
density ellipsoid is weakest near the virial radius \citep{WGK13}.
One would therefore expect that the departure of the stack and median $q(r)$
profiles would be greatest in the inner regions of haloes. Instead,
it would be worthwhile testing these ideas by stacking the haloes along their principal
axes before measuring $q$ in either circular or elliptical annuli,
but this is beyond the scope of this
work. In any event, the analysis of the randomly stacked halo is useful for
comparisons with observational studies that stack randomly (without aligning
first the principal axes) quasi-circular
astronomical systems (e.g.,
\citealp{McKay+02,Prada+03,Conroy+07_dyn,KP09,Wojtak&Mamon13} for probing SDSS galaxy haloes
with the kinematics of their satellites).

Cluster-mass haloes are currently merging, hence not very relaxed. On the other
hand, galaxy-mass haloes have assembled their mass much earlier and are more
relaxed at $z=0$. Could more relaxed regions lead to $q$ closer to unity?
While there is considerable scatter ($\sigma_q \approx 0.1$) between the
$q(r)$ profiles of individual haloes (Fig.~\ref{img_qofrall}), there is little
modulation with mass (Fig.~\ref{img_q_bestfit_median}), in the 1 dex
cluster-mass range studied here.

One should not over-interpret the results of our $q$-Gaussian fits to small
radii, such as $r = 0.03\, r_{100}$, close to the expected position of the Sun
in the Milky Way's halo. Indeed, 
the best-fit
$q$-Gaussian model strongly over-predicts the fraction of objects with
velocities greater than $4\,\sigma$ (Fig.~\ref{img_G_contours_some}). 
If the threshold for direct dark matter
detection is that high, then the best-fit $q$-Gaussian model will strongly
over-predict the dark matter detection rate.

It is interesting to compare our results at $r=0.03\,r_{100}$ to those
obtained by other workers.
\cite{LNAT10} found that the SDF of the dark matter in their hydrodynamical cosmological simulation was well fit by a
$q$-Gaussian with $q=0.70$ (after translating from their formula with
exponent $q/(1-q)$ to our exponent of $1/(1-q)$). Hence, their SDF is more truncated than the Gaussian, while our
SDF with $q=1.17$ (both stack and median) is less truncated than
a Gaussian. 
\cite{LSWW11} also find SDFs that are
more truncated at large velocities than the best-fit Gaussian (see their Fig.~3).
In comparison, the top left panel of Fig.~\ref{img_G_u_some}
also shows that the simulated SDF is flatter than the best-fit Gaussian
model, and falls off faster at large $u$. The preference for the $q=1.17$
Tsallis model comes from the low-end half of the SDF.
Our result is robust, as none of our 90 haloes has $q<1$ at $r =
0.03\,r_{100}$ (Fig.~\ref{img_qofrall}, although a half-dozen haloes have too
few particles for reliable determinations at this radius). Also, for our lower
mass haloes, this radius is only 1.5 times the softening length of the
cosmological simulation, so softening effects may play a role.

The differences between our results and those of previous authors might be
explained by the different behaviour of the radial and tangential VDFs:
at the small radius corresponding to the Solar radius in the Milky Way halo,
\cite{FS09} and \cite{Kuhlen_2010} find more truncated than Gaussian radial
VDFs but more extended than Gaussian tangential VDFs, as \cite{HS12} also found at
all radii where the density slope satisfied $\gamma \geq -2.4$.
Our anisotropic VDF (and our  SDF) involve
some averaging between the radial and tangential components:  (see
equation~\ref{eq_u_v}).  
It would be worthwhile to redo the analysis presented here using
equation~(\ref{eq_vdf_from_df}) on the separable DF that
\cite{Wojtak+08} measured on simulated $\Lambda$CDM haloes.

Finally, the analysis presented here indicates that it is not optimal to assume
Gaussian 3D velocities, as currently implemented in MAMPOSSt.
The inclusion of  our anisotropic $q$-Gaussian VDF 
(equation~\ref{eq_anisotropic_F_general}) into MAMPOSSt would improve the mass /
anisotropy modeling of this algorithm, for example using the shape of $q(r)$ of
equations~(\ref{eq_q_r1}) and (\ref{eq_q_r2}), and possibly forcing its parameters.

\section*{Acknowledgements}

We thank Ronan Lacire for checking equations~(\ref{eq_Dq}) and (\ref{eq_Cq}),
Charles Mazuet for preliminary work on the non-separability of VDFs, and Joe
Silk for useful discussions.
We also warmly thank the referee, Steen Hansen, for enlightening remarks.
L.B.eS. acknowledges support from the Brazilian CAPES foundation and from the
French CNRS through the GRAVASCO workshop.
R.W. acknowledges support through the Porat 
Postdoctoral Fellowship. The Dark Cosmology Centre is funded by the Danish National Research Foundation.
\bibliography{references_leandro,master}
\label{lastpage}
\end{document}